\documentclass{article}

\usepackage{comment}
\usepackage{hyperref}
\usepackage{natbib} 
\usepackage{siunitx}
\usepackage{float}
\usepackage{amsmath}
\usepackage{amssymb}
\usepackage{cleveref}
\usepackage{booktabs}

    \usepackage{arxiv}
    \title{Analytical modelling of wind-turbine wake turbulence in neutral atmospheric boundary layers}

    \newif\ifuniqueAffiliation
    \uniqueAffiliationtrue

    \author{ 
    \href{https://orcid.org/0000-0002-3252-2781}{\includegraphics[scale=0.06]{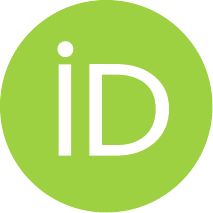}\hspace{1mm}
    Fr\'ed\'eric Blondel
    }\\
	IFP Energies nouvelles\\
    92852 Rueil-Malmaison, France \\
	\texttt{frederic.blondel@ifpen.fr} \\
	\And
	\href{https://orcid.org/0000-0001-5024-364X}{\includegraphics[scale=0.06]{figures/orcid.pdf}\hspace{1mm}Erwan J\'ez\'equel} \\
	IFP Energies nouvelles\\
    92852 Rueil-Malmaison, France \\
	\texttt{erwan.jezequel@ifpen.fr} \\
	\And
	\href{https://orcid.org/0000-0001-7504-2908}{\includegraphics[scale=0.06]{figures/orcid.pdf}\hspace{1mm}Helen Schottenhamml} \\
	IFP Energies nouvelles\\
    92852 Rueil-Malmaison, France \\
	\texttt{helen.schottenhamml@ifpen.fr} \\
	\And
	\href{https://orcid.org/0000-0001-7504-2908}{\includegraphics[scale=0.06]{figures/orcid.pdf}\hspace{1mm}Majid Bastankhah} \\
	Department of Engineering, \\
    Durham University, \\
    Durham DH1 3LE, UK\\
	\texttt{majid.bastankhah@durham.ac.uk} \\    
}    

\usepackage{xfrac}

\usepackage{siunitx}
\usepackage{cancel}
\usepackage[acronym]{glossaries}
\glsdisablehyper
\newacronym{abl}{ABL}{atmospheric boundary layer}
\newacronym{sbl}{SBL}{smooth boundary layer}
\newacronym{rbl}{RBL}{rough boundary layer}
\newacronym{tke}{TKE}{turbulent kinetic energy}
\newacronym{pde}{PDE}{partial differential equation}
\newacronym{les}{LES}{large eddy simulation}
\newacronym{rans}{RANS}{Reynolds-averaged Navier-Stokes}
\newacronym{lbm}{LBM}{lattice Boltzmann method}
\newacronym{pdf}{PDF}{particle distribution function}
\newacronym{ip}{IP}{isotropisation of production}
\newacronym{cwbl}{CWBL}{coupled wake boundary layer}

\newcommand{\altfrac}[2]{\ifmmode\def\tmp{$}\else\def\tmp{}\fi\mbox{%
    {\raisebox{.24\ht\strutbox}{\tmp#1\tmp}}%
    \kern-2.2pt\scalebox{1.6}[1.5]{/}\kern-1.8pt%
    {\tmp#2\tmp}%
    }}

\newcommand{\tr}[1]{%
  \ifmmode
    \textcolor{black}{#1}%
  \else
    \textcolor{black}{#1}%
  \fi
}

\newcommand{\revTwo}[1]{%
  \ifmmode
    \textcolor{black}{#1}%
  \else
    \textcolor{black}{#1}%
  \fi
}

\newcommand{\reynolds}[2]{\ensuremath{\overline{#1'#2'}}}
\newcommand{\up}{\reynolds{u}{u}}
\newcommand{\bu}{\overline{u}}
\newcommand{\tauKW}{\tau_{k,w}}

\DeclareDocumentCommand\advection{ o }{{\IfNoValueTF{#1}{\ensuremath{\mathcal{A}_{k}}}{\ensuremath{\mathcal{A}_{k, #1}}}}}

\DeclareDocumentCommand\dissipation{ o }{{\IfNoValueTF{#1}{\ensuremath{\mathcal{\varepsilon}_{k}}}{\ensuremath{\mathcal{\varepsilon}_{k, #1}}}}}

\DeclareDocumentCommand\production{ o }{{\IfNoValueTF{#1}{\ensuremath{\mathcal{P}_{k}}}{\ensuremath{\mathcal{P}_{k, #1}}}}}

\DeclareDocumentCommand\transport{ o }{{\IfNoValueTF{#1}{\ensuremath{\mathcal{T}_{k}}}{\ensuremath{\mathcal{T}_{k, #1}}}}}

\DeclareDocumentCommand\pressuretransport{ o }{{\IfNoValueTF{#1}{\ensuremath{\mathcal{D}_{k}}}{\ensuremath{\mathcal{D}_{k, #1}}}}}

\DeclareDocumentCommand\diffusion{ o }{{\IfNoValueTF{#1}{\ensuremath{\mathcal{V}_{k}}}{\ensuremath{\mathcal{V}_{k, #1}}}}}

\usepackage{tabularx}
\newcolumntype{C}{>{\centering\arraybackslash}X}

\begin{document}
\maketitle

\begin{abstract}
So-called engineering or analytical wind farm flow solvers typically build upon two submodels: one for the velocity deficit and one for the wake-added turbulence intensity. While velocity deficit modelling has received considerable attention, wake-added turbulence models are less prevalent in comparison. Yet, accurate estimates of local turbulence intensity are essential for predicting flow interactions and energy yield, as turbine wakes are both sensitive to, and sources of turbulence. Existing wake-added turbulence models are typically empirical or assume axial symmetry despite the inherently three-dimensional nature of turbulent wake fields. 
In this work, we present a \revTwo{new} model for wake-added turbulence intensity. Our approach is based on the analysis of the \acrlong{tke} and the streamwise Reynolds stress budget, incorporating classical \acrlong{rans} modelling assumptions and far-wake approximations. 
The resulting model maintains a simple and practical form, demonstrating strong agreement with \acrlongpl{les} and wind tunnel measurements. Our model provides a more physically consistent and predictive tool for wind farm flow modelling and performance estimation.
\end{abstract}

\section{Introduction}
\label{sec:intro}
Wind farm layout design or real-time control typically rely on simplified flow models, often based on analytical wake formulations. For yield assessment and power production modelling of finite-size wind farms, the common practice is to model cumulative wake effects using analytical wake-deficit models. A key parameter in these models is the wake recovery rate, which depends heavily on the local turbulence intensity at the \tr{wake-emitting rotor position}. The velocity-deficit models are thus often coupled with a wake-added turbulence model \citep[see][]{Niayifar2016,Bastankhah2021,Schmidt2023,Bay2023,pyWake2023,Blondel2023}.
While velocity deficit models \citep{Katic1987,Bastankhah2014,Keane2016,Blondel2020,Ishihara2018} have received significant attention, wake-added turbulence models have remained considerably less examined, despite their importance to overall flow solver accuracy.

The topic of axisymmetric wakes behind bluff bodies subjected to uniform inflows is extensively discussed in turbulence textbooks as a classic case of free-shear flows \citep[e.g.,][]{Tennekes1972,Pope2000}. Yet, wind turbines typically operate within sheared and turbulent atmospheric boundary layers. The change in flow conditions introduces additional complexity and leads to asymmetric wake shapes, particularly in the vertical profiles of wake turbulence \citep{Chamorro2008}, where even the interaction of turbines with the inflow boundary layer may reduce the turbulence level close to the ground \citep{Xie-Archer,bastankhah2017new}.

Wake-added turbulence models for wind turbines are mostly either empirical or provide limited information on the turbulence structure in turbine wake flows. \citet{Crespo1996}, for instance, approximated the streamwise evolution of turbulence in turbine wakes, neglecting any cross-flow spatial variations. The other model, which provides only a single representative value of the wake-added turbulence at each streamwise position, is the one developed by \citet{Frandsen1992}. It is derived based on the top-down approach of modelling wind farms \citep{Frandsen2006}.
More recent empirical three-dimensional models by \citet{Ishihara2018} or \citet{Khanjari2025} include shear effects via correction terms or fitted shape functions. Similarly, \citet{Stein2019} proposed axisymmetric analytical models of the three velocity variances based on wind tunnel measurements behind three-bladed wind turbines. \citet{Jezequel2024b} showed that the wake-added turbulence is sensitive to atmospheric stability and can be modelled by splitting turbine-induced and meandering-induced added turbulence. 

More recently, \citet{Bastankhah2024} proposed a model for axisymmetric wake flows. This model mathematically solves a simplified version of the \gls{tke} transport equation. In addition to assuming axisymmetry, which does not hold for turbine wakes in boundary layer flows, the model relies on numerical integration, making it less suitable for computationally intensive tasks such as wind farm layout optimisation.

In this work, we propose a new three-dimensional analytical model for the \gls{tke} and the streamwise normal Reynolds stress~\up{} in wind turbine wakes. Section~\ref{sec:wakeTurbulence} describes the datasets generated using a \gls{les} solver based on the \gls{lbm} for \tr{a range of ground roughness and wind turbine thrust coefficients}. Afterwards, we analyse the \gls{tke} transport equation budget in section~\ref{sec:simplifiedTkePde} and derive a simplified \gls{pde} for \gls{tke} transport that is solved numerically.
Using far-wake assumptions, we then propose an analytical model for the \gls{tke} written in an explicit and closed-form expression in section~\ref{sec:analyticalTKE}.
We extend the discussion in section~\ref{sec:upupExtension} to predictions of \up{}.
Lastly, we validate model predictions against wind tunnel measurements in section~\ref{sec:windTunnel}.

\section{\texorpdfstring{\acrshort{les}}{LES} wake-added \texorpdfstring{\acrshort{tke}}{TKE} and \texorpdfstring{\up{}}{streamwise Reynolds stress} evolution behind a porous disk in boundary layers}\label{sec:wakeTurbulence}
Throughout this article, we employ the following notations. Three axes, 
 $x_i$ for $i = 1, 2, 3$, span the Cartesian coordinate system of the domain. Here, $x_i$ corresponds to the streamwise axis $x$, the lateral axis $y$, or the vertical axis $z$, respectively. Similarly, $u_i$ denotes the corresponding velocity components $u$, $v$ and $w$. Temporally averaged quantities read $\overline{\{\cdot\}}$ and the deviation from the mean $\{\cdot\}'$. 
The Reynolds decomposition, which splits the instantaneous velocity $u_i$ into the mean velocity $\overline{u}_i$ and the velocity fluctuation $u_i'$, then reads $u_i = \overline{u}_i + u_i'$.  

\subsection{Large eddy simulation: the waLBerla-wind flow solver}\label{sec:waLBerla_wind_intro}

To establish and validate the developed analytical model, we use data obtained from the LES of porous disks immersed in boundary layers.  For this, we use waLBerla-wind, a solver based on the \gls{lbm}. Unlike traditional approaches, the \gls{lbm} does not aim to solve the Navier-Stokes equations for macroscopic quantities, such as pressure and velocity, but solves the lattice Boltzmann equation:
\begin{equation}
    f_q(x+c_q\Delta t, t + \Delta t) - f_q(x, t) = \Omega_q(x,t) + S_q(x, t).
    \label{eq:LBM}
\end{equation}
Here, $f_q$ are the \glspl{pdf}, which represent the density of particles at a given time~$t$ and position~$x$ with a specific discrete velocity~$c_q$, $\Omega_q$ is the collision operator that relaxes the \glspl{pdf} towards their equilibrium, and $S_q$ denotes a volumetric source term. The mass and momentum density follow from the \glspl{pdf} as~\mbox{$\rho = \sum_q f_q$} and~\mbox{$\rho u = \sum_q f_q c_q$}.

The discrete velocity set $\{c_{q}\}$ emerges during the discretisation of the continuous velocity space that adds to the temporal and spatial discretisation on a Cartesian grid. In this study, we use a three-dimensional velocity set with $27$ discrete velocities, represented by a $D3Q27$ stencil, which ensures high precision, stability, and rotational invariance \citep{Bauer2018}. 
Furthermore, we employ the cumulant \acrlong{lbm} \citep{Geier2015} along with a fourth-order correction \citep{Geier2017}. The QR minimum-dissipation subgrid-scale model \citep{Verstappen2011,Verstappen2014} accounts for unresolved eddies. For further details about waLBerla and its extension to wind turbines and atmospheric flows, see \citet{Bauer2021} and \citet{Schottenhamml2022, Schottenhamml2024}.

\tr{In the context of \revTwo{analytical} wake-added turbulence modelling, mainly purely empirical models have been introduced so far. Our primary goal is to propose an approach that more realistically captures the physics of wind turbine wakes and their interaction with the ABL. To achieve this, we deliberately adopted a simplified setup to first address the truly neutral situation, similar to, e.g., see~\citet{Wu2011,Stevens2018,MouLin2019}. Due to neglected Coriolis forcing, no wind veer is present in the flow. Atmospheric stratification is likewise neglected. Stratification is well known to affect wake structure and recovery, see, e.g.,~\citet{Hancock2014,Jezequel2024b}. Our approach accounts for how wake-added turbulence is impacted by ambient turbulence, which in turn is influenced by thermal stratification. However, the more complex impact of thermal stratification on the size of atmospheric turbulent structures, wake meandering, and related phenomena is neglected in this work for simplicity.}


The simulation setup is as follows: the domain has dimensions of $36D$ in length, \tr{$12D$} in width, and \tr{$5D$} in height, where $D=\SI{0.15}{\metre}$ is the diameter of the wind turbine. For simplicity, we model the turbine as a non-rotating, uniformly loaded actuator disk, as is often done in experimental \citep{Aubrun2013} and numerical studies \citep{Wu2011}. We follow \citet{Stevens2014} and impose a modified thrust coefficient $C_T^{'}=C_T/(1-a)^2$ with the thrust \tr{coefficients $C_T= \qtylist{0.4;0.6;0.8}{}$} and the wind turbine induction factor $a=1/2(1-\sqrt{1-C_T})$. The mean velocity at hub-height $z_h$ is $\overline{u}_h=\SI{2.5}{\metre\per\second}$. \tr{Appendix~\ref{app:lesValidation} shows the Reynolds number independence of our results, by reproducing the simulation with a utility scale rotor of diameter $D=\SI{150}{\m}$ and a reference wind velocity at hub height of $u_h= \SI{8}{\metre\per\second}$.}

Following \citet{Munters2016}, we apply shifted periodic boundary conditions in the streamwise direction to prevent the development of locked large-scale turbulent structures. A free-slip boundary condition at the top of the domain represents the finite height of the \gls{abl}, and a wall boundary condition \citep{Han2021} adapted to the $D3Q27$ stencil models the ground of the domain. The wall boundary condition calculates the shear stress based on the Monin-Obukhov similarity theory.
A concurrent methodology is applied, as described in \citet{Dhamankar2018}. We allow the \gls{abl} in the ``empty" simulation run to develop over \tr{$150$} flow-through times before starting the main simulation, which includes the wind turbine. Afterwards, both simulations run for an additional \tr{$300$} flow-through times. Data is collected during the last \tr{$150$} flow-through times to exclude wake build-up effects. The concurrent simulation provides the inflow plane for the main simulation, and a non-reflective outflow condition models the domain outlet. 

\tr{For a domain height of $\delta = 5D$, the ground roughness is varied over $z_0/\delta = \num{1e-7}, \num{1e-6}, \num{1e-5}, \num{1e-4}, \num{1e-3}$ in order to obtain time-averaged hub-height turbulence intensities $I_x = \sqrt{\up{}}/\bu_h$ between $5\%$ and $13\%$. This range represents typical offshore conditions, as well as moderately rough onshore conditions \citep{EWA_1989}. 
In addition, further simulations were performed using $z_0/\delta = \num{1e-7}$ and varying domain heights, $\delta = \num{3}D, \num{4}D, \num{6}D$. These cases are analysed in Section~\ref{subsec:inflowEddyViscosity}.}

\begin{figure}[!ht]
    \centering
    \includegraphics[width=10.5cm]{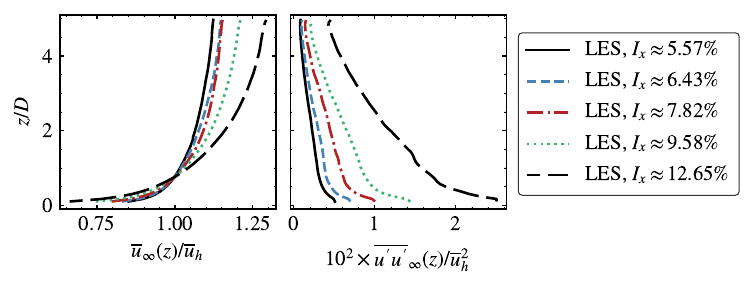}
    \caption{Normalised velocity (left) and streamwise normal Reynolds stress (right) inflow profiles for all considered roughness lengths.}
    \label{fig:inflowProfiles}
\end{figure}

\tr{
Figure~\ref{fig:inflowProfiles} presents the resulting inflow velocity profiles (i.e., unaffected by the wake) and streamwise normal Reynolds stress profiles extracted four rotor diameters upstream of the actuator disk and laterally aligned with the rotor centre. In the remainder of this manuscript, the subscript 
$\infty$ denotes such time-averaged vertical inflow profiles. The corresponding time-averaged hub-height turbulence intensities are reported in the figure legend.
}

\tr{Among these simulations, we will refer to the cases with $C_T=0.8$ and $z_0/\delta=\num{1e-7}$ as the \gls{sbl} case and $C_T=0.8$ and $z_0/\delta=\num{1e-4}$ as the \gls{rbl} case. For these cases, the turbulence intensities at hub height are approximately \qtylist{5.5;9.5}{\%}. The other cases are less extensively presented in the analysis, but are included in the model calibration routines. 
This LES setup will be validated against laboratory experiments later in section~\ref{sec:windTunnel}.}

\tr{In the remainder of this paper, the wake is divided into two regions: the near wake and the far wake. Following \citet{Bastankhah2016}, the near-wake length $x_0$ is estimated to range between $3$ and $5$ rotor diameters downstream of the disk, depending on the configuration. In this work, we adopt $x_0 = 4D$ as the transition between the near-wake and far-wake regions.}

\subsection{Velocity deficit, \texorpdfstring{\gls{tke}}{TKE} and \texorpdfstring{\up{}}{streamwise Reynolds stress} evolution}
\begin{figure}[!ht]
    \centering
    \includegraphics[width=12.cm]{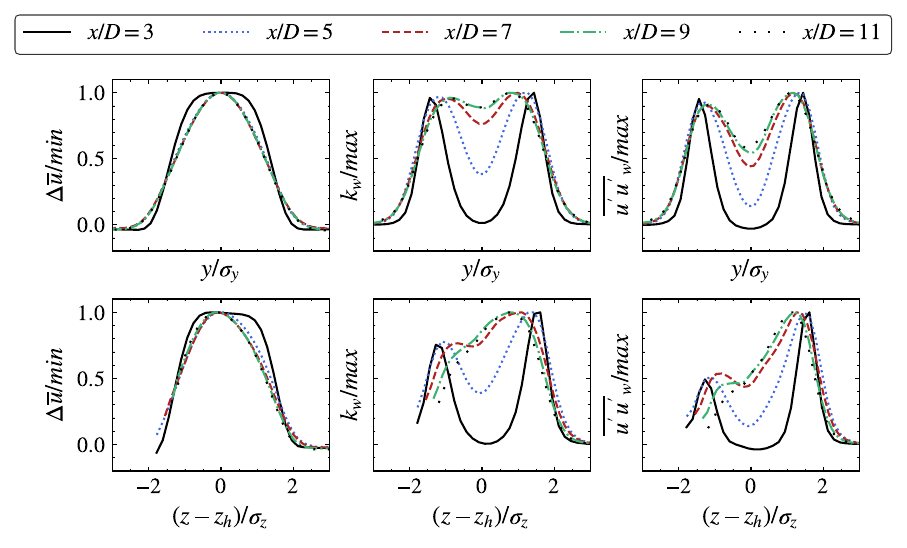}
    \caption{~\gls{sbl} case lateral and vertical profiles of the velocity deficit~$\Delta \bu{}$, the wake-added \gls{tke}~$k_w$ and the wake-added~$\up{}_w$ profiles in the wake of the porous disk in a smooth boundary layer, normalised by their minimum or maximum values.}
    \label{fig:selfSimilarity}
\end{figure}

Figure~\ref{fig:selfSimilarity} shows lateral and vertical profiles of velocity deficit~$\Delta \bu{}=\overline{u}-\overline{u}_\infty$, wake-added \gls{tke}~$k_w=k-k_\infty$ and wake-added streamwise turbulence $\up{}_w=\up{}-\up{}_\infty$ at different streamwise positions in the wake. \tr{The subscript $w$ represents the wake quantities, whereas the absence of a subscript stands for the total quantities}. At each streamwise position, the quantities are normalised by their minimum or maximum values; spatial dimensions by the wake width in the vertical and lateral directions, denoted by $\sigma_z$ and $\sigma_y$, respectively. 

Our simulations predict the same self-similar behaviour of the velocity deficit in the vertical and lateral directions as reported in experimental observations \citep{Bastankhah2014}. The wake-added \gls{tke} and $\up{}_w$, however, show a transition from a bi-modal distribution in the near-wake towards a more uniform distribution in the far wake, as confirmed by the budget analysis in section~\ref{sec:budget-wake-added-transport-equation}. True self-similarity occurs only in the very far wake \tr{(for $x>9D$)}, as also observed in the experiments of \citet{Stein2019}. 
Furthermore, strong three-dimensionality is observed: lateral and vertical profiles differ significantly. Consequently, the assumption of self-similarity \tr{is not deemed} justified for the development of an analytical model for wake-added turbulence. 

\noindent
Generally, the wake-added \gls{tke} and $\up{}_w$ show very similar trends. Under the hypotheses we will introduce in section~\ref{sec:pressureStrainModel}, both of them differ only in the presence of the pressure‐strain correlation term in the $\up{}_w$ transport equation \citep[see][among others]{Lumley1975}.
Hence, we will treat $\up{}_w$ as an extension of the wake-added \gls{tke} models later in this study.

\section{A simplified \texorpdfstring{\acrshort{tke}}{TKE} transport equation}\label{sec:simplifiedTkePde}
We begin by simplifying the wake-added \gls{tke} transport equation to derive a relatively simple \gls{pde}, which can be solved numerically using a one-dimensional marching-forward scheme in the streamwise direction. In the next section, this is further simplified in the far wake to develop an explicit analytical model for wake-added turbulence, eliminating the need for numerical schemes \tr{and reducing the computational time}.

\subsection{\texorpdfstring{\acrshort{tke}}{TKE} transport equation}\label{sec:TKE_budget}
For an incompressible fluid with no external forces, the time-averaged \gls{tke} transport equation is \citep{Pope2000}
\begin{equation}
    \label{eq:tkeBudget}
    \underbrace{\nu \overline{ \frac{\partial u'_i}{\partial x_j}\frac{\partial u'_i}{\partial x_j} }}_{\dissipation{}}
     = 
    \underbrace{-\bu{}_j\frac{\partial k}{\partial x_j}}_{\advection{}}\,
    \underbrace{-\frac{1}{\rho}\frac{\partial \overline{u'_i p'}}{\partial x_i}}_{\pressuretransport{}} \,
    \underbrace{-\frac12\frac{\partial \overline{u'_j u'_j u'_i}}{\partial x_i}}_{\transport{}}\,
    \underbrace{+\nu \frac{\partial^2 k}{\partial x_j^2}}_{\diffusion{}}\,
    \underbrace{-\reynolds{u_i}{u_j}\frac{\partial \bu{}_i}{\partial x_j}}_{\production{}}
    ,
\end{equation}
where $k=1/2\, \overline{u_i'u_i'}$. \advection{} is the advection by the mean flow, \pressuretransport{} is the transport by turbulent pressure fluctuations, \transport{} is the transport by turbulent velocity fluctuations, \diffusion{} is the viscous diffusion, where $\nu$ is the dynamic viscosity, \production{} is the shear production, and \dissipation{} is the \gls{tke} dissipation. 
\tr{The regularised cumulant~\gls{les} model that we use is a mixture of explicit/implicit~\gls{les}, as discussed in \citet{Gehrke2022,Spinelli2023}. 
In such a context, the~\gls{tke} dissipation is a combination of numerical dissipation, the implicit part, and the contribution of the subgrid-scale model (i.e., the energy transfer from the filtered velocity field to the sub-grid scale motions), which is the explicit part. We thus cannot compute the dissipation directly from the subgrid-scale model, as in \citet{Klemmer2024}. Instead, the dissipation is taken as the residual of equation~\eqref{eq:tkeBudget}, as in \citet{Heinze2015} or \citet{Colombie2021}. We also neglect the turbulent diffusion due to sub-grid stresses as their impact on the budget analysis is expected to be small~\citep{Klemmer2024}. 
}
Subtracting equation~\eqref{eq:tkeBudget}, the total \gls{tke} transport equation, from its counterpart based on the \tr{inflow} conditions yields the wake-added transport equation:
\begin{equation}
    \label{eq:tkeABudget}
    \underbrace{\dissipation{} - \dissipation[\infty]{}}_{\dissipation[w]{}}
     = 
    \underbrace{\advection{} - \advection[\infty]{}}_{\advection[w]{}} + 
    \underbrace{\pressuretransport{} - \pressuretransport[\infty]{}}_{\pressuretransport[w]{}} + 
    \underbrace{\transport{} - \transport[\infty]{}}_{\transport[w]{}} + 
    \underbrace{\diffusion{} - \diffusion[\infty]{}}_{\diffusion[w]{}} + 
    \underbrace{\production{} - \production[\infty]{}}_{\production[w]{}}
    .
\end{equation}

\subsection{Budget of the wake-added \texorpdfstring{\acrshort{tke}}{TKE} transport equation}\label{sec:budget-wake-added-transport-equation}

\begin{figure}[!ht]
    \centering
    \includegraphics[width=12cm]{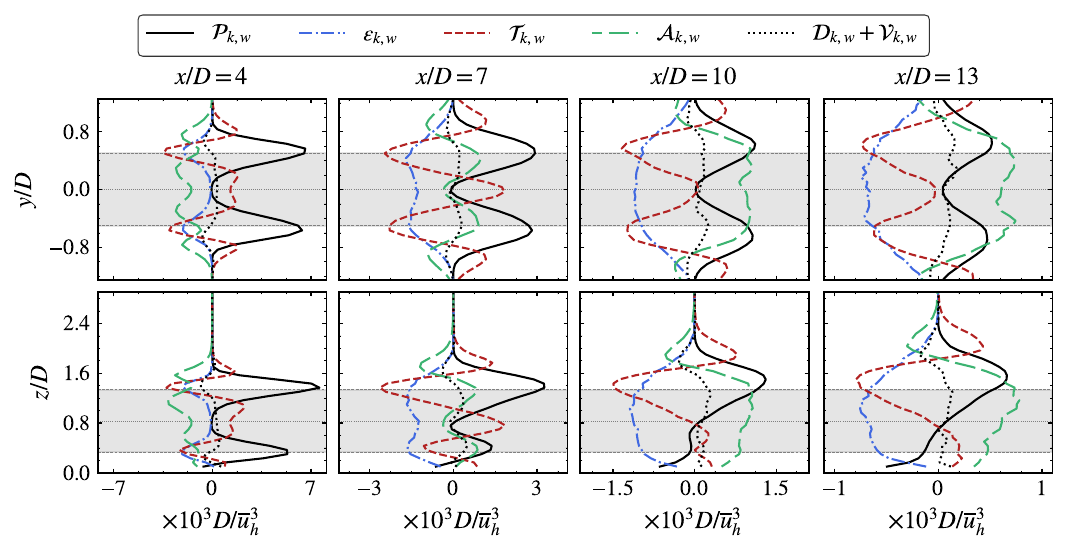}
    \caption{\tr{Normalised lateral (top) and vertical (bottom) profiles of the  wake-added \gls{tke} budget terms for the \gls{sbl} case.}}
    \label{fig:tkeBudgets}
\end{figure}

Figure~\ref{fig:tkeBudgets} shows the budgets of the wake-added \gls{tke} transport equation. The four dominant terms are the production~\production[w]{}, the advection~\advection[w]{}, the transport~\transport[w]{} and the dissipation~\dissipation[w]{}. The two remaining terms, the viscous diffusion~\diffusion[w]{} and the transport by turbulent pressure fluctuations~\pressuretransport[w]{} are \tr{deemed negligible: their maximum value is at most one fourth of the predominant terms and mostly below 10\% in the wake region.} 


\noindent
The production term in equation~\eqref{eq:tkeBudget} directly depends on the velocity gradients. 

The velocity gradients are high in the tip regions, leading to large values of production. 
However, below the hub height, the wake-generated gradients are typically negative, which counteracts the positive inflow gradient, leading to low or even negative added-turbulence production. Note that negative values of~\production[w]{} do not indicate a sink of \gls{tke} but rather a lower production compared to the upstream region. The turbulent transport term acts as a diffusion mechanism, redistributing \gls{tke} from high- to lower-magnitude regions. We can also observe that the \gls{tke} dissipation exhibits a bimodal shape, which evolves with streamwise distance into a more uniform shape. Lastly, it is worth noting that in the wake centre, the advection term initially acts as a sink in the near wake and later becomes a source in the far wake, as it advects the high levels of \gls{tke} from the near wake to the far wake.

\subsection{Term-by-term analysis}\label{sec:term-by_term}
According to the analysis of the wake-added \gls{tke} budget in the previous section, the simplified wake-added \gls{tke} equation consists of four terms, each of which we will now discuss and simplify for wind turbine wakes: 
\begin{equation}
    \label{eq:tkeABudget}
    \underbrace{\dissipation{} - \dissipation[\infty]{}}_{\dissipation[w]{}}
     = 
    \underbrace{\advection{} - \advection[\infty]{}}_{\advection[w]{}} + 
    \underbrace{\transport{} - \transport[\infty]{}}_{\transport[w]{}} + 
    \underbrace{\production{} - \production[\infty]{}}_{\production[w]{}}
    .
\end{equation}
In the wake, the streamwise component of the advection term, $-\bu{}\,\partial k / \partial x$, dominates over the other components, as shown in appendix~\ref{app:APT} (figure~\ref{fig:simpleAdvection}). Considering that $\partial k_\infty/\partial x = 0$, the wake-added advection simplifies to
\begin{equation}
    \mathcal{A}_{k,w}\approx -\bu{}\dfrac{\partial k}{\partial x}\approx -\bu{}\dfrac{\partial k_w}{\partial x}.
    \label{eq:tkeAdvection}
\end{equation}
Similar to the advection term, one can argue that the turbulent transport outside of the wake,~\transport[\infty]{}, is negligible, resulting in $\transport[w]{}\approx \transport{} = - 1/2\, \partial (\overline{u_i'u_i'u_j'}) / \partial x_j$.
We apply the gradient diffusion hypothesis \citep[see][]{Harlow1969,Jones1972,Pope2000} on the turbulent transport~\transport{}, and thus, approximate the triple velocity correlations in terms of the turbulent Prandtl number~$\text{Pr}_t$ (assumed to be one) and the eddy viscosity~$\nu_t$\tr{, defined later in section~\ref{sec:eddyViscVdLaan},} as follows:
\begin{equation}
    -\frac12\overline{u_i'u_i'u_j'}\approx \left(\nu+\frac{\nu_t}{\text{Pr}_t}\right)\frac{\partial k}{\partial x_j} \approx \frac{\nu_t}{\text{Pr}_t}\frac{\partial k}{\partial x_j},
    \label{eq:tripleCorrelations}
\end{equation}
where $\nu$ is the kinematic viscosity of the fluid.
%
\tr{Using equation~\eqref{eq:tripleCorrelations}, we can model the wake-added turbulent transport term as
\begin{equation}
    \mathcal{T}_{k,w}\approx \dfrac{\partial }{\partial x_j}\left( \frac{\nu_t}{\text{Pr}_t}\frac{\partial k}{\partial x_j}\right)\approx \dfrac{\partial }{\partial y}\left( \frac{\nu_t}{\text{Pr}_t}\frac{\partial k_w}{\partial y}\right)+\dfrac{\partial }{\partial z}\left( \frac{\nu_t}{\text{Pr}_t}\frac{\partial k_w}{\partial z}\right).
    \label{eq:tkeDiffusion}
\end{equation}}
The validity of these simplifications is further assessed in  figure \ref{fig:simpleDiffusion} in appendix~\ref{app:APT}.

For the wake-added production term~$\production[w]{} = \production{} - \production[\infty]{}$, we evaluate all Reynolds stresses using Boussinesq's eddy-viscosity hypothesis, $\reynolds{u_i}{u_j}=2/3k\delta_{ij}-2\nu_t S_{ij}$, defined in terms of the Kronecker delta $\delta_{ij}$ and the rate-of-strain tensor $S_{ij}=\frac12\left( \partial \bu{}_i/\partial x_j + \partial \bu{}_j/\partial x_i \right)$. 
Generally, only gradients of the streamwise mean velocities contribute significantly to the \gls{tke} production. Moreover, the terms $\up{}\,\partial \bu{}/\partial x$, $\partial \bu_\infty{}/\partial x$ and $\partial \bu_\infty{}/\partial y$ are negligible. The validity of these assumptions is again discussed in appendix~\ref{app:APT} (figure \ref{fig:simpleProduction}).
The wake-added production term is therefore simplified to
\begin{equation}
    \mathcal{P}_{k,w}= \nu_t\left[ 
    \left(\frac{\partial \bu{}}{\partial y}\right)^2+\left(\frac{\partial \bu{}}{\partial z}\right)^2 \right]
        -   
    \nu_{t,\infty}
    \left(\frac{\partial \bu_\infty}{\partial z}\right)^2
    .
    \label{eq:tkeProduction}
\end{equation}
\tr{Lastly, we derive the wake-induced dissipation term based on dimensional analysis in terms of wake-added \gls{tke} and a turbulent dissipation time scale~$\tauKW$. The time scale is defined as $\tauKW\propto 1/\omega$, i.e., proportional to the inverse of the specific dissipation rate used in the classical $k$–$\omega$ turbulence model developed in~\citet{Wilcox2008}. This choice is consistent with the $k$–$\tau$ turbulence modelling framework introduced in~\citet{Speziale1992}. The dissipation takes the form}
\begin{equation}
    \dissipation[w]=\dfrac{k_w}{\tauKW}.
    \label{eq:tkeDissipation}
\end{equation}
\noindent
\tr{An alternative approach would be to model the dissipation as: $\dissipation[w] = c k_w^{3/2}/{l_m}$ } \citep{Pope2000}, where \( l_m \) is the mixing length scale and \( c \) is a model constant . However, we avoid this formulation for two main reasons: first, we prefer working with a linear equation for \( k_w \)\tr{, as it later greatly facilitates the derivation of an analytical model}. Second, our \gls{les} data indicate that \( \tauKW \) takes a simpler three-dimensional form in the wake (shown later in section \ref{sec:dissipation-time-scale-models}) than a mixing-length-based model, where both $l_m$ and $c$ can potentially vary in the wake, particularly for non-equilibrium turbulent flows~\citep{Bastankhah2024}. \tr{Appendix~\ref{app:TauEvolution} supports these hypotheses, showing comparisons of the turbulent time scale and turbulent length scale behaviours within wakes at different thrust coefficients and ground roughness.}


Combining equations~\eqref{eq:tkeAdvection}-\eqref{eq:tkeDissipation} results in a simplified \gls{pde} based on the eddy viscosities~$\nu_t$ and~$\nu_{t,\infty}$, turbulent dissipation time scale~$\tauKW$, and the mean velocity field: 
\begin{equation}
\tr{
    \bu{}\dfrac{\partial k_w}{\partial x}=\nu_t\left[ 
    \left(\frac{\partial \bu{}}{\partial y}\right)^2+\left(\frac{\partial \bu{}}{\partial z}\right)^2 \right]
        -   
    \nu_{t,\infty}
    \left(\frac{\partial \bu_\infty}{\partial z}\right)^2 
        +
        \dfrac{\partial }{\partial y}\left( \frac{\nu_t}{\text{Pr}_t}\frac{\partial k_w}{\partial y}\right)+\dfrac{\partial }{\partial z}\left( \frac{\nu_t}{\text{Pr}_t}\frac{\partial k_w}{\partial z}\right)
    -\dfrac{k_w}{\tauKW}.
    }
    \label{eq:simplifiedPDE}
\end{equation}
The mean velocity field can be taken directly from experimental data, numerical simulations, or an analytical wake model. In the latter case, it is worth noting that since solving equation~\eqref{eq:simplifiedPDE} includes a forward-marching scheme in the streamwise direction, any inaccuracy in near-wake modelling \tr{may} propagate and affect far-wake predictions. Therefore, using an engineering model that provides realistic near-wake predictions, such as those proposed by \citet{Blondel2020} and \citet{schreiber2020brief}, is recommended. The dependence of far-wake predictions on near-wake properties is mitigated in Section~\ref{sec:analyticalTKE}, where we develop an analytical TKE model for the far wake.

Assuming a given velocity field and \tr{inflow conditions}, equation~\eqref{eq:simplifiedPDE} is closed by estimating the eddy viscosity $\nu_t$ and the dissipation time scale $\tauKW$, which are elaborated in the following sections.

\subsection{Estimation of turbulent viscosity \texorpdfstring{$\nu_t$}{}}\label{sec:eddyViscVdLaan}
\begin{figure}[!ht]
    \centering
    \includegraphics[width=\textwidth]{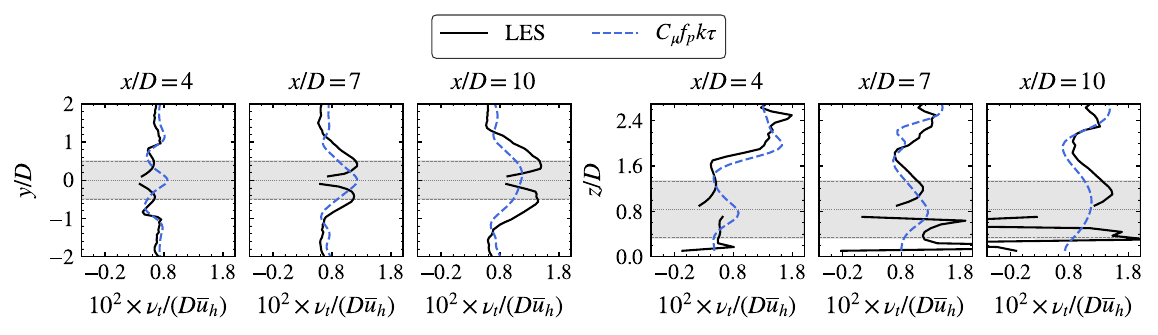}
    \caption{Comparison of the modelled and \gls{les}-based lateral and vertical wake eddy viscosity profiles inside the wake for the \gls{sbl} case.}
    \label{fig:eddyViscosities}
\end{figure}

\tr{Based on dimensional analysis, the eddy viscosity can be expressed as the product of the turbulent kinetic energy,~$k$, and a turbulent time scale, $\tau_k$. In the classical formulation of~\citet{Launder1974}, the eddy viscosity is written as 
\begin{equation}
    \nu_t=C_\mu \dfrac{k^2}{\varepsilon_k},
\end{equation}
with the model constant $C_\mu$. Introducing the turbulent time scale $\tau_k=k/\varepsilon_k$, this expression can be equivalently reformulated as
\begin{equation}
    \nu_t=C_\mu k \tau_k.
\end{equation}
As noted by~\citet{vanDerLaan2015}, this approach yields a model that is too dissipative in the near wake, where the velocity gradients are high. To address this, we adopt the method proposed by \citet{vanDerLaan2015} and introduce a flow-dependent coefficient $C_\mu^*=f_p C_\mu$. The scalar function $f_p$ models the effect of non-equilibrium conditions, and reads:
\begin{equation}
    \label{eq:fpFunction}
    f_p=\dfrac{2f_0}{
            1+\sqrt{1+4f_0\left(f_0-1\right)
                \left(
                    \dfrac{\sigma}{\tilde{\sigma}}
                \right)^2
            }
    }.
\end{equation}
Here, $\sigma$ is the so-called shear parameter:
\begin{equation}
    \sigma=\frac{k}{\varepsilon_k}\sqrt{\left(\frac{\partial u_i}{\partial x_j}\right)^2}=\tau_k\sqrt{\left(\frac{\partial u_i}{\partial x_j}\right)^2}\approx \tau_k\sqrt{\left(\frac{\partial u}{\partial y}\right)^2+\left(\frac{\partial u}{\partial z}\right)^2}.
\end{equation}
which is given by $\widetilde{\sigma}=1/\sqrt{C_\mu}$ outside of the wake. $f_0$ is a constant factor, expressed in terms of the Rotta constant: $f_0=C_R/\left(C_R-1\right)$.
Note that an alternative approach was recently proposed in~\citet{Klemmer2025}, consisting in applying the inflow eddy viscosity in the near-wake region, not used here.
Figure~\ref{fig:eddyViscosities} shows the performance of this approach, using $C_\mu=0.06$ and $C_R=2.5$. For the~\gls{les} data, the eddy viscosity is computed from
$
    \nu_t = {\mathcal{P}_k}/\left({ 
    \left({\partial \bu{}}/{\partial y}\right)^2+\left({\partial \bu{}}/{\partial z}\right)^2}\right)$.
}

\tr{
The figure shows that the overall trends are captured by the model. At $x/D=4$, the model predicts a total eddy viscosity lower than $\nu_{t,\infty}$ in the wake region. In the lateral direction, at larger streamwise distances, the model tends to underpredict the eddy viscosity. In the vertical direction, the overall agreement is satisfactory, despite large oscillations in the \gls{les} data in regions with small velocity gradients, where a near-zero denominator amplifies the uncertainty in estimating the turbulent eddy viscosity.}


\tr{
To prevent the generation of spurious \gls{tke} in the undisturbed flow, the eddy viscosity outside the wake region is constrained to its ambient value. To this end, we introduce a wake sensor based on velocity gradients, defined as}
\begin{equation}
    \left(\dfrac{\partial u}{\partial y}\right)^2 + \left(\dfrac{\partial u}{\partial z}\right)^2 - \left(\dfrac{\partial u_\infty}{\partial y}\right)^2 > \epsilon_{\text{wake}}.
\end{equation}
\tr{
Outside the region identified by this criterion, we enforce $\nu_t = \nu_{t,\infty}$. In practice, a threshold value of $\epsilon_{\text{wake}} = 10^{-4}$ is used.
}

\subsection{Estimation of dissipation time scale \texorpdfstring{$\tauKW$}{}}\label{sec:dissipation-time-scale-models}

To model the wake-added dissipation time scale, $\tauKW$, we start with the definition of the wake-added dissipation $\dissipation[w]{}$,
\begin{equation}
    \dissipation[w]{} = \dfrac{k_w}{\tauKW}=\dissipation-\dissipation[\infty]=\dfrac{k}{\tau_k}-\dfrac{k_\infty}{\tau_{k,\infty}}.
    \label{eq:wakeDissipation}
\end{equation}
Quantifying $\tauKW$ directly based on the \gls{les} data is challenging since near the wake edges, where both $k_w$ and $\varepsilon_w$ are small, computing $\tauKW = k_w / \varepsilon_w$ is \tr{subject} to numerical errors. Instead, we use the \gls{les} data to estimate $\tau_k$ and $\tau_{k,\infty}$ and assess which one is more representative of $\tauKW$ at different locations. Figure~\ref{fig:Tau} shows vertical and lateral variations of $\tau_k$ obtained from the \gls{les} data at two different streamwise locations, where $\tau_{k,\infty}$ is also shown as a reference. \tr{Additional results are presented in Appendix}~\cref{app:TauEvolution}\tr{, where the evolution of $\tau_{k,\infty}$ is presented along lateral and vertical profiles across a range of thrust coefficients and streamwise distances.} Figure \ref{fig:Tau} shows that in the lateral direction, the value of $\tau_k$ remains almost constant in the wake and increases to $\tau_{k,\infty}$ outside the wake. 
\tr{
\begin{figure}[!ht]
    \centering
    \includegraphics[width=13.5cm]{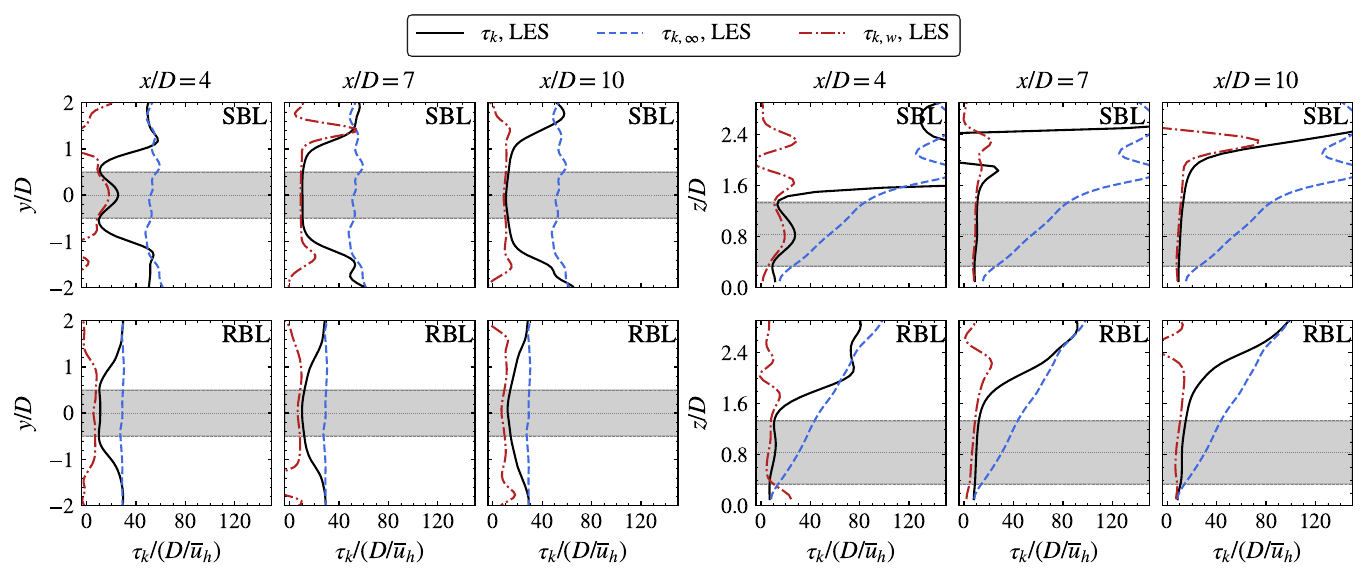}    
    \caption{\tr{Lateral and vertical profiles of the dissipation time scale for the SBL (top) and RBL (bottom) cases at two different downwind locations. Reference results at $x=-3D$ are shown as blue dashed lines.}}
    \label{fig:Tau}
\end{figure}
}
%
In the vertical profiles, we have very different behaviour in each of the three regions. Below the wake, we have $\tau_k \approx \tau_{k,\infty}$, a behaviour particularly evident in the \gls{rbl} case. Therefore, from equation~\eqref{eq:wakeDissipation}, we have below the wake 
\begin{equation}
\frac{k_w}{\tauKW} \approx \frac{k}{\tau_{k,\infty}}-\frac{k_\infty}{\tau_{k,\infty}}= \frac{k_w}{\tau_{k,\infty}},
\end{equation}
resulting in $\tauKW \approx \tau_{k,\infty}$. 
Inside the wake, both $k>k_{\infty}$ and $\tau_k<\tau_{k,\infty}$ hold; hence, 
\begin{equation}
    k_w / \tauKW \approx k / \tau_k, \quad \text{and} \quad \tauKW \approx \tau_k (1-k_\infty/k) \approx \tau_k.
\end{equation}
Lastly, above the wake, the wake-added \gls{tke} vanishes quickly, and therefore, the choice of $\tauKW$ becomes less significant. For simplicity, we assume $\tauKW \approx \tau_k$ above the wake.

\begin{figure}[!ht]
    \centering
    \includegraphics[width=9.cm]{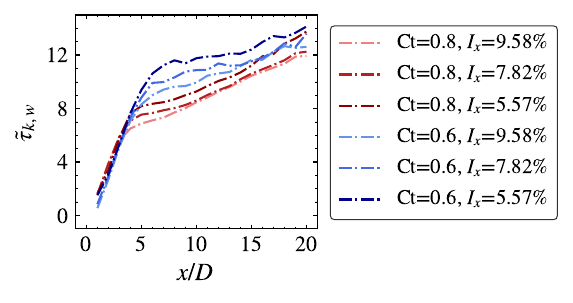}    
    \caption{\tr{Evolution of the dissipation time scale $\tilde{\tau}_{k}$ within the wake for different ground roughness and thrust coefficients.}}
    \label{fig:TauC_Evo}
\end{figure}

Using $z_{BT}$ as the height of the wind turbine bottom-tip, the dissipation time scale results as
\begin{equation}\label{eq:tau_w}
    \tauKW= 
    \begin{cases}
        \tau_{k,\infty} & \text{if } z\leq z_{BT}=z_h-D/2, \\
        \tau_k              & \text{if } z>z_{BT}.
    \end{cases}
\end{equation}
To use equation \eqref{eq:tau_w}, we still need to model  $\tau_{k,\infty}$ and $\tau_k$.
Figure~\ref{fig:TauC_Evo} shows the streamwise evolution of \tr{$\tilde{\tau}_{k,w}$} \tr{for a selected number of~\gls{les}}, where the tilde denotes \tr{the best fit, obtained from a least-square minimisation procedure between the model and the~\gls{les} data.} A linear fit seems to provide a good approximation of $\tilde{\tau}_{k,w}$ in the far wake region. Interestingly, the slopes of the linear fits seem to be \tr{independent of the thrust coefficient and turbulence intensities, whereas the y-intercepts differ slightly depending on the flow and actuator disk conditions}.
Therefore, $\tilde{\tau}_{k}$ is assumed to change linearly with $x$ within the wake region according to
\begin{equation}
    \tilde{\tau}_{k} = a_\tau x + b_\tau,
\end{equation}
where $a_\tau$ and $b_\tau$ are constants. \tr{Numerical experiments indicate that treating $\tau_{k,w}$ as independent of the thrust coefficient yields satisfactory results while preserving the simplicity of the formulation. In the remainder of this article, we use $a_\tau=0.33$ and $b_\tau=8.1-33.4I_x$, where $I_x$ is the streamwise turbulence intensity at hub height outside the wake. The validity of this approximation has been assessed over the following ranges of thrust coefficients and turbulence intensities: $0.4 \le C_T \le 0.8$ and $0.05 \lesssim I_x \lesssim 0.12$.}

For the boundary-layer inflow, the two dominant terms in the TKE transport equation are expected to be the turbulent production and dissipation. Equating these two terms at the inflow gives $\tau_{k,\infty}=k_\infty/(\nu_{t,\infty}(\partial \bu_\infty/\partial z)^2)$. Assuming a logarithmic velocity profile, $\bu_\infty=(u_*/\kappa) \textrm{log}(z/z_0)$, where $u_*$ is the friction velocity, $\tau_{k,\infty}$ takes the form of
\begin{equation}
    \tau_{k,\infty} = \frac{k_\infty}{l_{m,\infty}^2}\left( \frac{\kappa}{u_*}z \right)^3.
    \label{eq:tauInfinity}
\end{equation}
Equation \eqref{eq:tauInfinity} can be simplified by assuming that, at low heights ($z < z_{BT}$), variations of $k_{\infty}$ with $z$ are small and that $l_{m,\infty} = \kappa z$. This implies $\tau_{k,\infty} \propto z$. For practical applications, where $k_\infty(z)$ is unknown, and to ensure a smooth transition at $z_{BT}=z_h-D/2$ for the values of $\tau_{k,w}$ given in equation \eqref{eq:tau_w}, a linear dependence of $\tau_{k,w}$ on $z$ for $0<z<z_{BT}$ can be defined as:
\begin{equation}
    \tauKW= 
    \begin{cases}
        \dfrac{z}{z_{BT}}\left( a_\tau x + b_\tau \right),& \text{if } 0<z\leq z_{BT} \\
        a_\tau x + b_\tau,              & \text{if }  z>z_{BT}.
    \end{cases}
    \label{eq:practicalTau}
\end{equation}

\subsection{Comparison of $k-(\tau)$ model with \texorpdfstring{\gls{les}}{LES} data \tr{and $k-(l)$ model}}

Before making further assumptions and simplifications to develop a fully closed analytical model, we evaluate the performance of the simplified \gls{tke} equation \eqref{eq:simplifiedPDE} against the \gls{les} data. \tr{This approach will be referred to as the $k-(\tau)$ model, $k$ referring to the resolved~\gls{tke} and the parentheses indicating the analytically prescribed closure variable}. Our aim here is solely to test the validity of the assumptions made in this section for deriving the simplified \gls{tke} equation, rather than to construct a complete model. Therefore, the LES data provide the velocity distribution and inflow eddy
viscosity, while equation~\eqref{eq:practicalTau} provides a linear fit of the
dissipation time scale; all quantities are then used in
equation~\eqref{eq:simplifiedPDE} to compute the wake-added \gls{tke}.
The advection term of the simplified \gls{pde} is discretized using a forward finite difference scheme, while the radial derivatives of velocity and wake-added \gls{tke} are computed using a central finite difference scheme. \tr{The wake-added \gls{tke} is initialised to zero at $x = 0$, and the second-order derivatives at the $y-$ and $z-$boundaries are estimated using a one-sided first-order scheme.}

\begin{figure}[!ht]
    \centering
        \includegraphics[width=13.5cm]{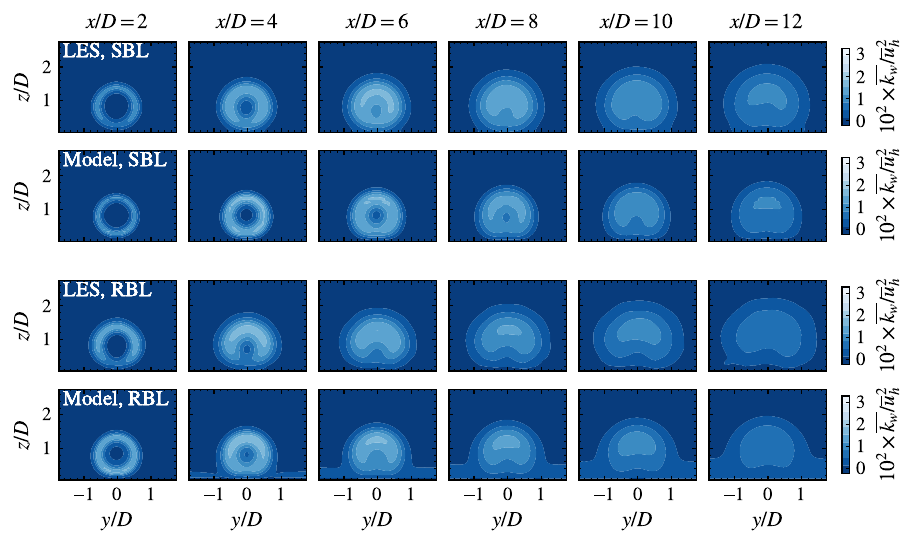}

    \caption{\tr{Contour plot of the wake-added \gls{tke}: constant $x$ planes at several streamwise distances. SBL case (first row: \gls{les}, second row: $k-(\tau)$ model) and the RBL case (third row: \gls{les}, fourth row: $k-(\tau)$ model) from the calibration dataset.}}

    \label{fig:GridModelContours}
\end{figure} 
\tr{Figure~\ref{fig:GridModelContours} compares the modelled wake-added \gls{tke} based on equation~\eqref{eq:simplifiedPDE} and the LES wake-added TKE.} 
It shows that, for both inflow conditions, the simplified numerical solution \tr{reproduces the~\gls{les} trends in the entire wake, including the near-wake region}. 
\tr{The overall wake shape and the three-dimensional structure of the wake-added \gls{tke} are well captured.}
\begin{figure}[!ht]
    \centering
    \includegraphics[width=13.7cm]{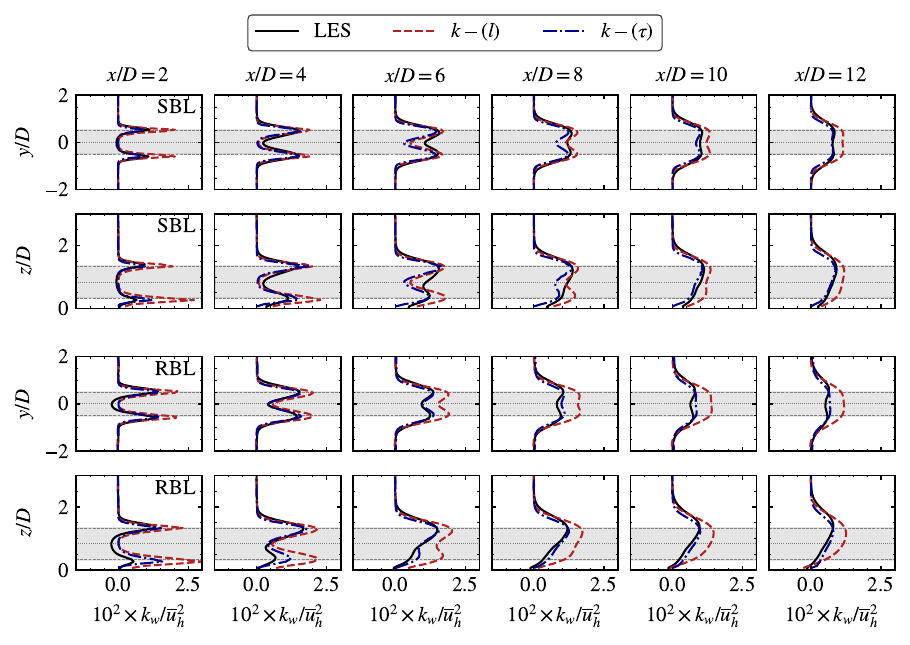}
    
    \caption{\tr{Lateral and vertical profiles of the wake-added turbulent kinetic energy based on the proposed $k-(\tau)$ model and the $k-(l)$ model, SBL case (first row: lateral profiles, second row: vertical profiles) and RBL case (third row: lateral profiles, fourth row: vertical profiles) from the calibration dataset.}}
    \label{fig:kL_vs_kTau_SBL}
\end{figure}

For a more quantitative analysis, lateral and vertical~\gls{tke} profiles predicted by the proposed $k-(\tau)$ model are compared with~\gls{les} data, and the $k-(l)$ model proposed in~\cite{Klemmer2025}. Here, $(l)$ indicates the analytically prescribed closure length scale used in ~\cite{Klemmer2025}. 
\revTwo{The parameters suggested in~\cite{Klemmer2025} are adopted for comparison, whereas the $k-(\tau)$ model uses values directly from the LES. In both cases, figure~\ref{fig:kL_vs_kTau_SBL}, the agreement between the $k-(\tau)$ model and the~\gls{les} data is very good, with only an under-estimation of the~\gls{tke} near the wake centre at $x/D=6$ and $x/D=8$. Despite being calibrated on a different dataset, the $k-(l)$ model also performs well. It captures the relative trends, but tends to overestimate the~\gls{tke}, in particular in the~\gls{rbl} case. As a conclusion, this section shows that equation \eqref{eq:simplifiedPDE} is a good approximation of the LES and that an analytical model can be built upon it.}

%

\section{An analytical model for the wake-added \texorpdfstring{\acrshort{tke}}{TKE}}\label{sec:analyticalTKE}
To develop an analytical solution for equation~\eqref{eq:simplifiedPDE}, we use far-wake assumptions to simplify the advection and transport terms further and obtain an explicit  algebraic formulation of $k_w$ rather than a \gls{pde}. Besides the simplified formulation, solving an algebraic relation for the far-wake region has the advantage that inaccuracies in the near wake are not propagated downstream and do not negatively affect far-wake predictions.

\subsection{Advection and transport terms simplification}        
\noindent
\tr{As shown in Appendix~\ref{app:APT} (figure~\ref{fig:simpleAdvection}), the advection term can be reduced to its streamwise component. Furthermore, we write the advection velocity as a streamwise varying velocity given by $\upsilon(x) \bu_h $, so:
\begin{equation}\label{eq:advSimplified}
    \mathcal{A}_{k,w} = -\bu\frac{\partial k_w}{\partial x} = -\upsilon(x)\bu_h\frac{\partial k_w}{\partial x}.
\end{equation}
}
\tr{Later in Section~\ref{sec:absorbingTKE},  $\upsilon(x)$ will be absorbed in a single, streamwise dependent constant $\mathcal{Z}_{k,w}(x)$, so hereafter $\upsilon(x)$ is dropped from the advection term for brevity.} To simplify the advection term (and only for this term), we assume self-similarity of the \gls{tke}. While it does not hold in the near and intermediate wake regions, self-similarity works reasonably well in the far wake (see figure~\ref{fig:selfSimilarity}), \tr{where the advection term is prominent (see figure~\ref{fig:tkeBudgets})}. 
In the context of self-similarity, we express the wake-added \gls{tke} as the product of the maximum value of the wake-added \gls{tke}, $K(x)$, and a shape function $g(\xi)$, i.e., $k_w=K(x)g(\xi)$. Here, $\xi=r/\sigma$ is the radial position $r$ normalised by the wake width $\sigma$ assuming wake axisymmetry, $\sigma=\sigma_y=\sigma_z$. With this, we obtain
\begin{equation}
    \mathcal{A}_{k,w}\approx  \underbrace{-\bu_h{}\frac{\mathrm{d}K(x)}{\mathrm{d}x}g(\xi)}_{\mathcal{A}_{k,w}^a}+\underbrace{\bu_h{}\frac{K(x)}{\sigma}\xi \frac{\mathrm{d}\sigma}{\mathrm{d}x}\frac{\mathrm{d}g}{\mathrm{d}\xi}}_{\mathcal{A}_{k,w}^b}.
    \label{eq:selfS_Advection1}
\end{equation}
%
%
Assuming that the maximum wake-added \gls{tke} evolves as $K(x)=a_K(x-x_0)^\alpha$ for a virtual origin $x_0$ and constants $a_K$ and $\alpha$, we write the first term of equation~\eqref{eq:selfS_Advection1}, $\mathcal{A}_{k,w}^a$, as
\begin{equation}\label{eq:A_wa}
    \mathcal{A}_{k,w}^a = -\bu_h \frac{\mathrm{d}K(x)}{\mathrm{d}x}g(\xi) = -\bu_h \frac{\mathrm{d}K(x)}{\mathrm{d}x}\frac{1}{K(x)}k_w = -\bu_h \dfrac{\alpha}{(x - x_0)}k_w. 
\end{equation}
%

Using equations \eqref{eq:selfS_Advection1} and \eqref{eq:A_wa} the simplified advection term, then, reads   
\begin{equation}
     \mathcal{A}_{k,w} \approx - \bu_h \dfrac{\alpha}{x-x_0}k_w + \mathcal{A}_{k,w}^b.
     \label{eq:selfS_Advection2}
\end{equation}
As will be shown hereafter, $\mathcal{A}_{k,w}^b$ will not be explicitly modelled and will be absorbed in the transport term.

\begin{figure}[!ht]
    \centering
    \includegraphics[width=13.5cm]{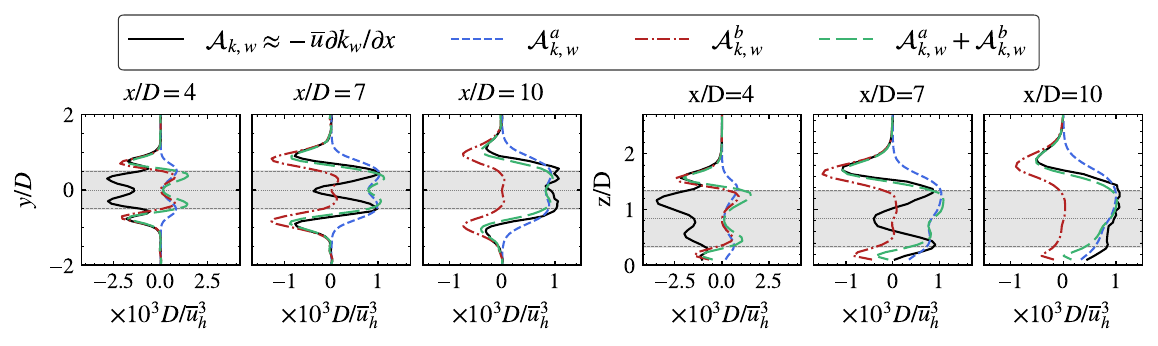}
    \caption{Lateral and vertical profiles of the advection term including $\mathcal{A}_{k,w}^a$ and $\mathcal{A}_{k,w}^b$ contributions for the~\gls{sbl} case}
    \label{fig:tkeAdvectionTerm}
\end{figure}

Figure~\ref{fig:tkeAdvectionTerm} compares the modelled advection term, derived using the above discussion, with the original term obtained from the LES data. 
The self-similarity-based approach provides a reasonable estimation of the advection term in the \tr{very} far wake, e.g., \tr{$x=10D$}, but performs less satisfactorily at shorter downwind distances, e.g., \tr{$x=7D$}, due to the breakdown of the self-similarity assumption closer to the turbine. 
\tr{
This assumption is considered acceptable because the contribution of the advection term to the budget is primarily significant in the far wake (i.e., for 
$x/D>7$; see figure~\ref{fig:tkeBudgets}) and is much less important in the near- and intermediate-wake regions. Improving the formulation of the advection term remains an area for future work, as equation~\eqref{eq:selfS_Advection2} relies on strong underlying assumptions.
}

Next, we simplify the transport term by analysing its behaviour across three regions of the wake cross-section: the wake core (central region), the intermediate region where shear is maximised, and the wake boundary (edge of the wake). 

\begin{figure}[!ht]
    \centering
   \includegraphics[width=8.5cm]{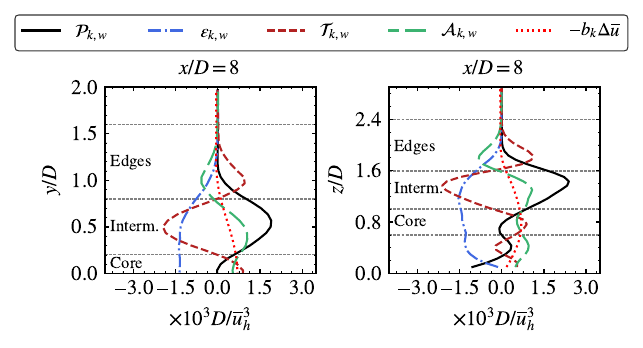}
    \caption{Lateral and vertical profiles of the diffusion term and its three constituting components. $c$ is an arbitrary constant, assumed negative here. The~\gls{sbl} case is considered here.}
    \label{fig:diffusionSimplification}
\end{figure}

Figure~\ref{fig:diffusionSimplification} shows the wake-added production~\production[w]{}, dissipation~\dissipation[w]{}, transport~\transport[w]{} and advection~\advection[w]{} at $x/D=8$. The transport and advection show similar behaviour at the wake edges. 
In this region, dissipation cancels production, i.e., $\dissipation[w]{}\approx -\production[w]{}$, leaving advection to cancel transport, as suggested by \citet{Tennekes1972}. According to figure~\ref{fig:tkeAdvectionTerm}, $\mathcal{A}_{k,w}^a$ is small in this region and we therefore assume that $\mathcal{A}_{k,w}^b$ cancels the transport, i.e., $\transport[w]{}\approx -\mathcal{A}_{k,w}^b$.

Within the intermediate region, where wake shear reaches its maximum, turbulence production is the dominant term. As shown in figure~\ref{fig:diffusionSimplification}, the transport term follows the distribution of production by acting as a sink of turbulence and transferring the generated turbulence to the neighbouring regions, i.e., wake centre and wake edge. We therefore assume that, in the intermediate region, the transport term can be modelled as  $\transport[w]{}\approx -a_k\production[w]{}$, where $a_k$ is a positive constant. Finally, in the wake core region, the transport term is the primary source of \gls{tke}, since the production term vanishes as $\partial \bu / \partial r$ approaches zero. We therefore model the transport term as being proportional to the velocity deficit, which is maximum in this region, i.e., $\transport[w]{} \approx -b_k \Delta \bu$, where $b_k$ is a positive constant.\\  
To model the transport term across all three regions, we represent it as the sum of the modelled contributions from each region. This linear superposition is justified by the fact that the contribution of each modelled term becomes small outside its respective region. For example, $\Delta \bu$ diminishes outside the wake core, production contributes mainly within the intermediate region, and $\mathcal{A}_{k,w}^b$ is close to zero except near the wake edges. The final equation describing transport across all regions
, therefore, reads
\begin{equation}\label{eq:transport_modelled}
    \transport[w]{}\approx -\mathcal{A}_{k,w}^b - a_k \production[w]{} - b_k\Delta \bu.
\end{equation}
Substituting the advection and the diffusion terms in equation~\eqref{eq:simplifiedPDE} with those derived in this section yields
\begin{equation}
        k_w \left(\bu_h\dfrac{\alpha}{x-x_0}+\dfrac{1}{\tauKW}\right) = 
        \nu_t\left(1-a_k\right)\left[\left(\dfrac{\partial \bu}{\partial y}\right)^2 +\left(\dfrac{\partial \bu}{\partial z}\right)^2 
        -  \dfrac{\nu_{t,\infty}}{\nu_t} \left(\dfrac{\partial \bu_\infty}{\partial z}\right)^2\right]        
        - b_k\Delta \bu.
    \label{eq:analyticalModel1}
\end{equation}

\subsection{Eddy viscosity ratios simplification}
At this point, equation~\eqref{eq:analyticalModel1} is an algebraic expression for the added \gls{tke} that can be solved to find $k_w$. However, it requires the calibration of six different parameters, which is cumbersome and increases the risk of overfitting the model. Experimental evidence \citep{Bastankhah2024} and numerical studies \citep{Iungo2017} have shown that $\nu_t$ approaches its asymptotic value $\nu_{t,\infty}$ in the far wake. For convenience, we therefore replace~$\nu_{t}$ in equation \eqref{eq:analyticalModel1} with~$\nu_{t,\infty}$ in the remainder of this article. It is worth noting that this assumption was not justifiable in the previous section, where equation~\eqref{eq:simplifiedPDE} was solved for the entire wake. Since we develop an algebraic equation applicable only to the far wake, the assumption is expected not to introduce substantial errors. \tr{We emphasise that this assumption is not mandatory in order to derive an algebraic model from equation \eqref{eq:analyticalModel1} but allows to minimise the number of model parameters.}

\begin{figure}[!ht]
    \centering
    \includegraphics[width=13.7cm]{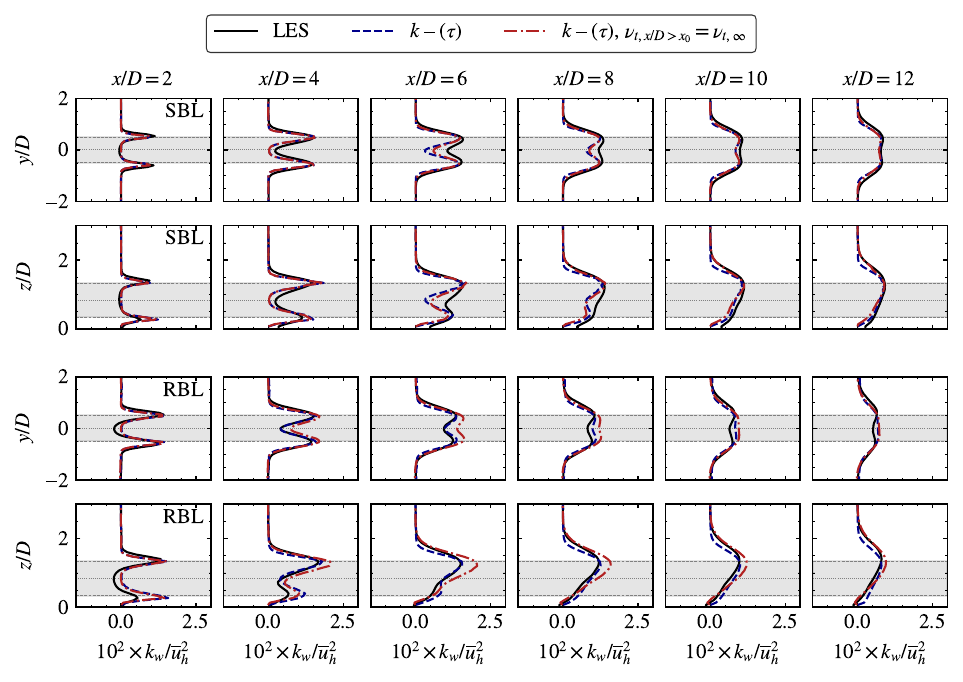}

    \caption{\tr{Lateral and vertical profiles of the wake-added turbulent kinetic energy based on the original $k-(\tau)$ model and a simplified eddy-viscosity formulation of it ($k-(\tau)$, $\nu_{t}=a_{\nu_t}\nu_{t,\infty}$), SBL case (first row: lateral profiles, second row: vertical profiles) and RBL case (third row: lateral profiles, fourth row: vertical profiles) from the calibration dataset.}}   
    \label{fig:simpleNuT_impact_SBL}
\end{figure}

\tr{In the following, the $k-(\tau)$ model is used to evaluate the impact of this hypothesis by enforcing $\nu_t=\nu_{t,\infty}$ in the far-wake region. In the near-wake region, the model is used as previously described in order to provide a realistic initial condition to the simplified model and focus on the far wake. The near-wake length $x_0$ is computed as introduced in~\citet{Bastankhah2016}. Downstream of $x_0$, we consider $\nu_t=\nu_{t,\infty}$. Results are shown in figure~\ref{fig:simpleNuT_impact_SBL}.
The impact on the results is considered acceptable for an analytical model, despite a noticeable overestimation of the wake-added ~\gls{tke} in the \gls{rbl} case.
}

\subsection{Grouping the advection and transport terms}\label{sec:absorbingTKE}
A final simplification consists in grouping the advection term~$\mathcal{A}_{k,w}^a \approx \bu_h \alpha/(x-x_0)$, the term $(1-a_k)$, and the inverse \tr{ of the } dissipation time scale~$1/\tauKW$ on the left-hand side of equation~\eqref{eq:analyticalModel1} as
\begin{equation}
    \dfrac{1}{\zeta_{k,w}} := \dfrac{1}{1-a_k}\left(\bu_h \dfrac{\alpha}{x-x_0} + \dfrac{1}{\tauKW}\right),
\end{equation}
and the parameters $b_k$ and $(1-a_k)$ on the right-hand side of equation~\eqref{eq:analyticalModel1} as
\begin{equation}
    \gamma_k = \dfrac{b_k}{1-a_k}.
\end{equation}

\tr{The streamwise evolutions of $\zeta_{k,w}$ and $\gamma_k$ are currently left undetermined, and will be studied in section~\ref{subsec:TKE_calibration}. Below the wake, where $z \leq z_{BT}$, we apply a linear decay on $\zeta_{k,w}$ towards zero at the wall, as we did for the dissipation time scale. $\zeta_{k,w}$ thus consist of a streamwise component, $\mathcal{Z}_{k,w}(x)$, that is damped near the ground:} 
%
\begin{equation}
    \zeta_{k,w}= 
    \begin{cases}
        \dfrac{z}{z_{BT}} \mathcal{Z}_{k,w}(x), &  \text{if } 0<z\leq z_{BT}  \\
        \mathcal{Z}_{k,w}(x),              &  \text{if }  z>z_{BT}.
    \end{cases}
    \label{eq:zeta}
\end{equation}

%
Using the term groupings introduced in this section and taking the limit $\nu_t \to \nu_{t,\infty}$, equation \eqref{eq:analyticalModel1} can be further simplified to develop the final form of the analytical model:
\begin{equation}
    \dfrac{k_w}{\zeta_{k,w}} = 
    \nu_{t,\infty}\left[\left(\dfrac{\partial \bu}{\partial y}\right)^2 +\left(\dfrac{\partial \bu}{\partial z}\right)^2 
    -  \left(\dfrac{\partial \bu_\infty}{\partial z}\right)^2\right]        
    - \gamma_k\Delta \bu.
    \label{eq:analyticalModelFinal}
\end{equation} 
\tr{$\zeta_{k,w}$, $\gamma_k$ and $\nu_{t,\infty}$ are the model's parameters and will be defined in the next sections}.

\subsection{\tr{Model calibration}}\label{subsec:TKE_calibration}

\subsubsection{\tr{Inflow eddy viscosity}}\label{subsec:inflowEddyViscosity}
\tr{
A common choice for the inflow eddy viscosity is $\nu_{t,\infty} = l_{m,\infty}^2 \times(\partial \bu_\infty / \partial z)$, where $l_{m,\infty}$ is the mixing length in the ABL, and it is estimated based on the model proposed by~\citet{Blackadar1962}. This leads to 
}
\begin{equation}
\nu_{t,\infty} = \left( \frac{\kappa z}{1 + \kappa z / l_{m,\infty}^{\text{max}}} \right)^2 \frac{\partial \bu_\infty}{\partial z},
\label{eq:infiniteEddyViscosity}
\end{equation}
\tr{
where $l_{m,\infty}^{\text{max}}$ is the maximum mixing length in the atmosphere.} 
\begin{figure}[!ht]
    \centering
    \includegraphics[height=0.215\textwidth]{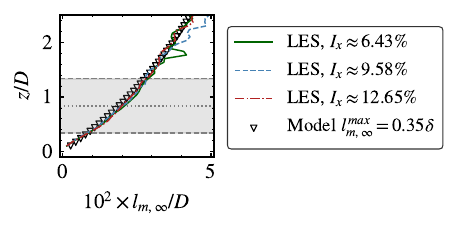} 
    \includegraphics[height=0.215\textwidth]{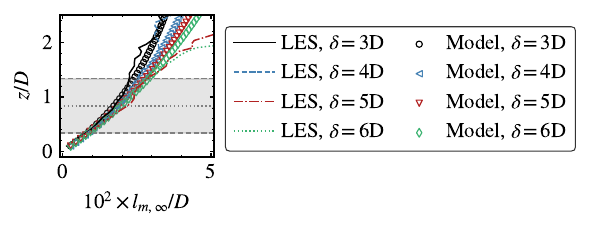}
    
    \caption{\tr{Vertical evolution of the inflow mixing length $l_m$ across three different ground surface roughness values for a domain height of $\delta=5D$ (left) and as a function of the domain height at a given ground roughness (right). $l_{m,\infty}^{ \text{max} }=0.35\delta$ is used in the model.}}    
    \label{fig:nutInflowEvolution_1}
\end{figure}
\tr{Consistent with previous studies \citep{Apsley1997,Tomas2011,Rodier2017}, we set $l_{m,\infty}^{\text{max}} = 0.35\delta$, where the boundary-layer thickness $\delta$ is assumed to be comparable to the domain height.} 
\tr{Note that because the simulation does not include a capping inversion, the boundary layer grows with less resistance and is expected to cover the entire vertical extent of the domain.}
\tr{
Figure~\ref{fig:nutInflowEvolution_1} (left) shows that the model provides a reasonable estimate of the mixing length for different values of $I_x$ and shear at $\delta = 5D$.
}
\tr{
As shown in figure~\ref{fig:nutInflowEvolution_1} (right), the approximation $l_{m,\infty}^{\text{max}} = 0.35\delta$ provides consistent results for domain heights ranging from $\delta = 3D$ to $\delta = 6D$ at a fixed roughness length. The figure also indicates that, within the rotor region, the length scale profiles become independent of the domain height once the condition $\delta > 4D$ is satisfied.
}

\subsubsection{\tr{Modelling $\zeta_{k,w}$ and $\gamma_k$}}\label{subsec:zeta}
\begin{figure}[!ht]
    \centering

    \includegraphics[width=0.95\textwidth]{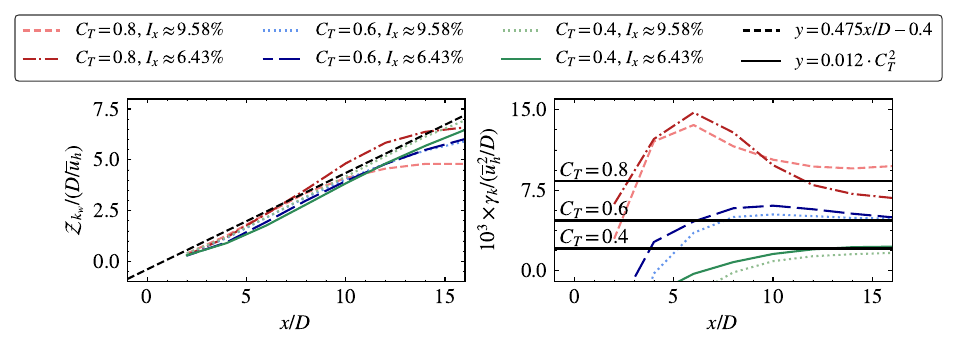}
    
    \caption{\tr{Streamwise evolution of $\mathcal{Z}_{k,w}$ (left) and $\gamma_k$ (right) for different thrust coefficients and inflow turbulence intensities, shown alongside the proposed linear fit.}}
    \label{fig:zetaEvolution}
\end{figure}

\tr{
The evolution of $\zeta_{k,w}$ and, subsequently, $\gamma_k$, is computed with a best-fit procedure between the \gls{les} data and the model at several streamwise locations
downstream of the turbine. The best fit minimises the root mean square error between the
model predictions and the \gls{les} results over each $y$-$z$ plane. As shown in figure~\ref{fig:zetaEvolution}, the parameter $\mathcal{Z}_{k,w}(x)=\text{max}(\zeta_{k,w}(x,z))$ is deemed independent of both the inflow turbulence intensity and the turbine thrust coefficient, and $\mathcal{Z}_{k,w} / (D/u_h) = 0.475\, x/D -0.4$ provides a good fit for the data in the far wake region.} 

%

\tr{The evolution of $\gamma_k$, shown in figure~\ref{fig:zetaEvolution}, exhibits a strong dependence on the thrust coefficient and a weaker dependence on the inflow turbulence intensity. 
}
\tr{
Figure~\ref{fig:zetaEvolution} shows that $\gamma_k$ increases in the near-wake until it reaches an asymptotic value in the far wake region. To keep the formulation simple and reduce the number of tuning parameters, streamwise variations of $\gamma_k$ are neglected. Due to the predominance of the production term in the region where $\gamma$ is not constant, more advanced formulations that were tested (but not shown here) yielded only marginal improvements in the model's predictive accuracy. Additionally, $\gamma_k$ is assumed independent of the inflow turbulence intensity. Based on the model calibration, the following simple empirical relationship is proposed:
\begin{equation}
\gamma_k / (\bu_h^2/D) = 1.2 \times 10^{-2}\, C_T^2.
\end{equation}
}
    
\subsection{Comparison to \texorpdfstring{\gls{les}}{LES} data}

\begin{figure}[!ht]
    \centering
    \includegraphics[width=14cm]{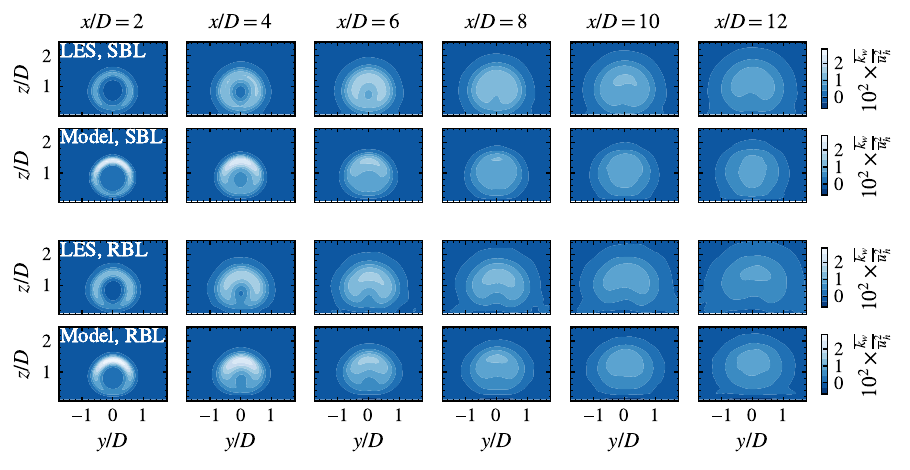}

    \caption{\tr{Contour plot of the wake-added \gls{tke}: constant $x$ planes at several streamwise distances. SBL case (first row: \gls{les}, second row: analytical model) and the RBL case (third row: \gls{les}, fourth row: analytical model) from the calibration dataset.}}    
    \label{fig:AnalyticalModelContours}
\end{figure}

Figure~\ref{fig:AnalyticalModelContours} compares the \tr{wake-added} \gls{tke} predicted using the proposed model, equation~\eqref{eq:analyticalModelFinal}, and the \gls{les} data. Here, the velocity fields from the \gls{les} simulations are used as input for the analytical model.

\begin{figure}[!ht]
    \centering
    \includegraphics[width=13.7cm]{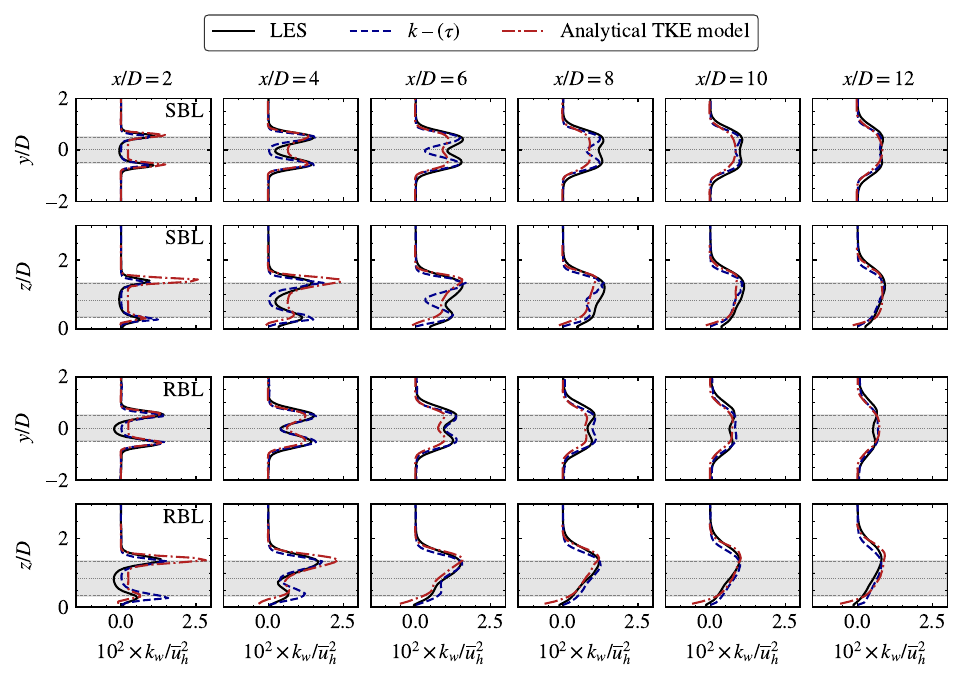}
    \caption{\tr{Lateral and vertical profiles of the wake-added turbulent kinetic energy. \gls{les} simulation,  $k-(\tau)$ model and analytical model predictions are compared for the SBL case (first row: lateral profiles, second row: vertical profiles) and the RBL case (third row: lateral profiles, fourth row: vertical profiles) from the calibration dataset.}}    
    \label{fig:AnalyticalModelProfiles_sbl}
\end{figure}

The model captures the wake shape correctly, showing a horseshoe shape in the near wake that smoothly transitions to a more uniform \gls{tke} distribution in the far wake. While the agreement in the far wake is good, we observe an overestimation of the \gls{tke} in the near-wake at $\tr{x/D=2}$ and $\tr{x/D=4}$, particularly near the top tip of the wake. 
This is to be expected, as the assumptions for this model are only valid in the far wake. 
\tr{A quantitative comparison is shown in figure~\ref{fig:AnalyticalModelProfiles_sbl}, where the lateral and vertical profiles of the added TKE are plotted. The analytical model tends to overpredict the~\gls{tke} near the wake edges in the near wake, i.e., at $x/D\le 4$. Further downstream, the agreement in both lateral and vertical directions is good. Small deviations are observed, but the overall trends are well captured.}
    

\section{Extension to Streamwise Reynolds normal stress \texorpdfstring{\up{}}{the streamwise Reynolds stress}}\label{sec:upupExtension}
Analytical wake velocity deficit models used in the wind energy industry \citep[e.g.,][among others]{Bastankhah2014,pena2016application, Blondel2020} typically rely on the streamwise turbulence intensity, instead of \gls{tke}, to estimate the wake expansion rate. This section extends the work of sections~\ref{sec:simplifiedTkePde} and~\ref{sec:analyticalTKE} to the wake-added normal turbulent stress, $\up{}_w$.

\subsection{Budget of the streamwise turbulent stress}

\begin{figure}[!ht]
    \centering
    \includegraphics[width=12cm]{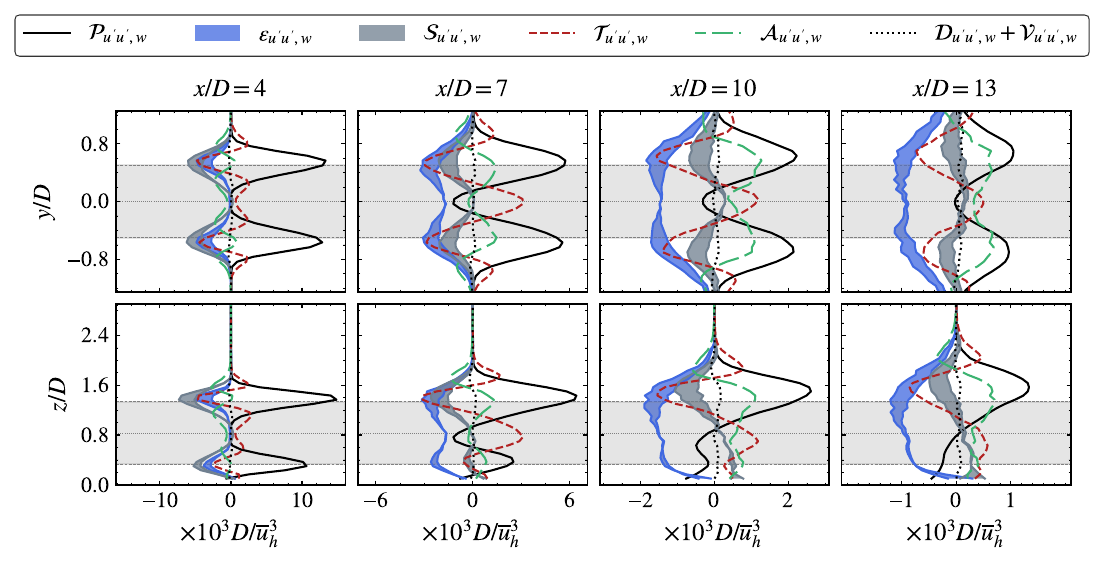}
    \caption{\tr{Horizontal and vertical profiles of the \up{} transport equation budget for the SBL case, with a range of potential solutions for the diffusion and pressure-strain correlations.}}
    \label{fig:upupBudget}
\end{figure}

The transport equation for the normal turbulent stress, under the same hypotheses as for the \gls{tke}, is \citep{Pope2000, Stull1988}:
\begin{equation}
     \underbrace{2\nu \overline{ \frac{\partial u'}{\partial x_i}\frac{\partial u'}{\partial x_i} }}_{\mathcal{\varepsilon}_{u'u'}} = - \underbrace{\bu{}_i\frac{\partial \up{}}{\partial x_i}}_{\mathcal{A}_{u'u'}} -
    \underbrace{
        \frac{\partial  \overline{u' u' u'_i}}{\partial x_i}}_{\mathcal{T}_{u'u'}} - 
    \underbrace{\frac{2}{\rho}\frac{\partial  \overline{p'u'}}{\partial x}}_{\mathcal{D}_{u'u'}} +
    \underbrace{
        \nu \frac{\partial^2 \up{}}{\partial x_i^2}
    }_{\mathcal{V}_{u'u'}}
    -
    \underbrace{2\overline{u' u'_i}\frac{\partial \bu{}}{x_i}}_{\mathcal{P}_{u'u'}} +
    \underbrace{2\overline{ \frac{p'}{\rho}\left( \frac{\partial u'}{\partial x}\right) }}_{\mathcal{S}_{u'u'}}
    .
    \label{eq:upupBudget}
\end{equation}
The normal turbulent stress transport equation has the same structure as the \gls{tke} transport equation~\eqref{eq:tkeBudget}, but with an additional term, the pressure-strain correlation term~$\mathcal{S}_{u'u'}$. 
Our \gls{les} flow solver may struggle to assess this term accurately as the SGS term may have a non-negligible contribution to it. 
\tr{Thus, two terms are assumed to be unknown in equation~\eqref{eq:upupBudget}: the dissipation term~$\mathcal{\varepsilon}_{u'u'}$ (due to the explicit/implicit nature of our LBM \gls{les} solver) and the pressure-strain correlation~$\mathcal{S}_{u'u'}$ (due to the contribution of SGS terms not captured by LES).} To determine all terms in equation \eqref{eq:upupBudget} from our LES data, we model the dissipation term and take the pressure-strain correlation as the residual of equation~\eqref{eq:upupBudget}. The dissipation tensor in anisotropic turbulence and the relation between different terms such as $\varepsilon_{u'u'}$ with $\varepsilon_k$ is an active area of research \citep[See][among others]{Perot2004, Gerolymos2016}. A common approach \citep{Hanjalic1976} is to blend an isotropic assumption, $\varepsilon_{u'u'}=2/3\varepsilon_k$, with Rotta's approach \citep{Rotta1951}, where the anisotropy of the dissipation tensor is assumed to be aligned with the anisotropy of the Reynolds stresses, i.e., $\varepsilon_{u'u'}=\varepsilon_k\up{}/k$. Rotta's approach is considered valid only at low Reynolds numbers, such as close to walls. Let $f$ represent the blending parameter, which can take values between $0$ and $1$. 
The dissipation writes 
\begin{equation}\label{eq:blendRottaIso} 
    \varepsilon_{u'u'}=(1-f)\frac{2}{3}\varepsilon_k + f\varepsilon_k \frac{\up{}}{k},
\end{equation}
where $\varepsilon_k$ is obtained from the \gls{tke} budget analysis discussed in section \ref{sec:TKE_budget}.
We consider the two limiting cases, $f=0$ and $f=1$, and, therefore, the enclosed range of possible solutions for $\varepsilon_{u'u'}$ and $\mathcal{S}_{u'u'}$. Figure~\ref{fig:upupBudget} shows the resulting budget at different streamwise positions.

The pressure-strain correlation~$\mathcal{S}_{u'u'}$ is part of the dominant terms. Since its role is to redistribute turbulence between the Reynolds stress components, it is a sink for the normal stress in most regions of the wake, except near the ground. Because the shear production of the other normal stresses $\reynolds{v}{v}$ and $\reynolds{w}{w}$ is negligible, they are primarily sustained by the redistribution of energy from the normal turbulent stress $\up{}$. 
Consequently, the pressure-strain correlation is sometimes referred to as a ``return to isotropy" term \citep[see][among others]{Rotta1951, Sarkar1990, Speziale1991}. 

\subsection{Pressure-strain correlation model} \label{sec:pressureStrainModel}
Under the hypothesis used in section~\ref{sec:simplifiedTkePde} to derive the \gls{tke} model, the \up{} production, diffusion and dissipation terms take the same simplified form as their \gls{tke} counterparts. Only the additional pressure-strain correlation term requires further investigation.  
Following \citet{Pope2000}, the pressure-strain takes the following form:
\begin{equation}\label{eq:Scrambling}
    \mathcal{S}_{u'_{i}u'_{j}}=\underbrace{-
    s_1\frac{\epsilon}{k}\left(\reynolds{u_i}{u_j}-\frac{2}{3}k\delta_{ij}\right)
    }_{\text{Rotta's slow term}}    
    \underbrace{-s_2\left(\mathcal{P}_{u'_i u'_j}-\frac{2}{3}\mathcal{P}_{k}\delta_{ij}\right)}_{\text{Rapid \gls{ip} term}}.
\end{equation}
Equation~\eqref{eq:Scrambling} is a linear combination of two models proposed by \citet{Launder1975}. The first term stems from Rotta's model \citep{Rotta1951} and accounts for the slow pressure effects. 
The second term is the so-called \acrfull{ip}~\citep{Naot1973} and represents the rapid pressure effects. The \gls{ip} counteracts the production of \up{}. Values of $s_1=1.8$ and $s_2=0.6$ are suggested in the literature \citep{Pope2000}. 

\begin{figure}[!ht]
    \centering
    \includegraphics[width=14cm]{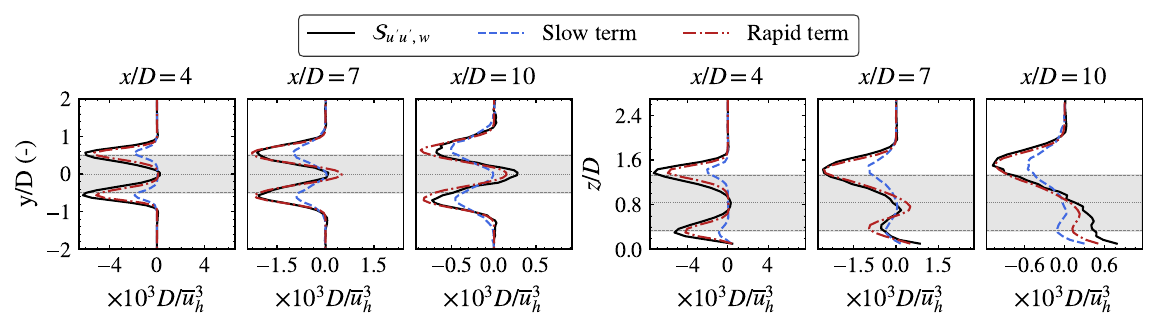}
    \caption{Comparison of the normalised components of the pressure strain term for the \gls{sbl} case. Hub-height horizontal profiles are shown to the left, and vertical profiles are shown to the right.}
    \label{fig:scamblingTerms}
\end{figure}

Figure~\ref{fig:scamblingTerms} shows the evolution of the two terms, formulated in terms of wake-added quantities, i.e., $\mathcal{S}_{u'u',w}=\mathcal{S}_{u'u'}-\mathcal{S}_{u'u',\infty}$, in the horizontal and vertical directions for the smooth boundary layer with low surface roughness. The overall behaviour is similar for the rough boundary layer (not shown here).
Both terms seem to vary similarly in the lateral and vertical directions, but the rapid term dominates roughly by a factor of two. Based on these observations, only the rapid term is retained in the following, with a modified value of the $s_2$ constant, accounting for the absorption of the slow term into the rapid term. Assuming that the production of the vertical and lateral stresses is negligible, i.e., $\mathcal{P}_{k,w} \approx \mathcal{P}_{u'u',w}$, the pressure-strain correlation term takes the simplified form
\begin{equation}\label{eq:ScramblingModel}
    \mathcal{S}_{u'u',w}= - s_3 \mathcal{P}_{u'u',w},
\end{equation}
where $s_3$ is a constant. 

\subsection{Simplified PDE and analytical model}

Using the model equations for advection, transport, production, dissipation and pressure-strain correlation, the simplified \gls{pde} for the normal Reynolds shear stress reads:

\begin{align}
\begin{split}
    \bu{}\dfrac{\partial \up{}_w}{\partial x} =& \tr{2}\nu_t \left(1-s_3\right)\left[ 
        \left(\frac{\partial \bu{}}{\partial y}\right)^2+\left(\frac{\partial \bu{}}{\partial z}\right)^2
    -   
        \dfrac{\nu_{t,\infty}}{\nu_t}
        \left(\frac{\partial u_\infty}{\partial z}\right)^2 
    \right]
     \\
        &
        \dfrac{\partial }{\partial y}\left( \frac{\nu_t}{\text{Pr}_t}\frac{\partial \up{}_w}{\partial y}\right)+\dfrac{\partial }{\partial z}\left( \frac{\nu_t}{\text{Pr}_t}\frac{\partial \up{}_w}{\partial z}\right)
        -\dfrac{\up{}_w}{\tau_{\up{},w}}.
    \label{eq:upupSimplifiedPDE}
    \end{split}
\end{align}

\tr{As in the $k-(\tau)$ model, it is possible to numerically solve equation~\ref{eq:upupSimplifiedPDE}. However, for the sake of brevity, we focus on the derivation of a closed-form expression for $\up{}$.} Following the same procedure used for the \gls{tke} model in section~\ref{sec:absorbingTKE} and integrating the newly introduced constant $s_3$ \tr{as well as the factor $2$ on the production term} into a function~$\zeta_{\up{},w}$ and a parameter~$\gamma_{\up{}}$, the analytical model reads
    \begin{equation}
        \dfrac{\up{}_w}{\zeta_{\up{},w}} = 
        \nu_{t,\infty}\left[\left(\dfrac{\partial \bu}{\partial y}\right)^2 +\left(\dfrac{\partial \bu}{\partial z}\right)^2 
        -  \left(\dfrac{\partial \bu_\infty}{\partial z}\right)^2\right]        
        - \gamma_{\up{}}\Delta \bu.
        \label{eq:upupAnalyticalModelFinal}
    \end{equation}
Therefore, the analytical models for the \gls{tke} and the streamwise turbulent stress are identical. They only differ in the chosen values for the parameters $\gamma_{\up{}}$ and $\zeta_{\up{},w}$. \tr{We remind here that we write:
\begin{equation}
    \zeta_{\up{},w}= 
    \begin{cases}
        \dfrac{z}{z_{BT}} \mathcal{Z}_{\up{},w}(x), &  \text{if } 0<z\leq z_{BT}  \\
        \mathcal{Z}_{\up{},w}(x),              &  \text{if }  z>z_{BT}.
    \end{cases}
    \label{eq:zeta_upup}
\end{equation}}

\subsection{\tr{Model calibration}}
\begin{figure}[!ht]
    \centering
    \includegraphics[width=0.95\textwidth]{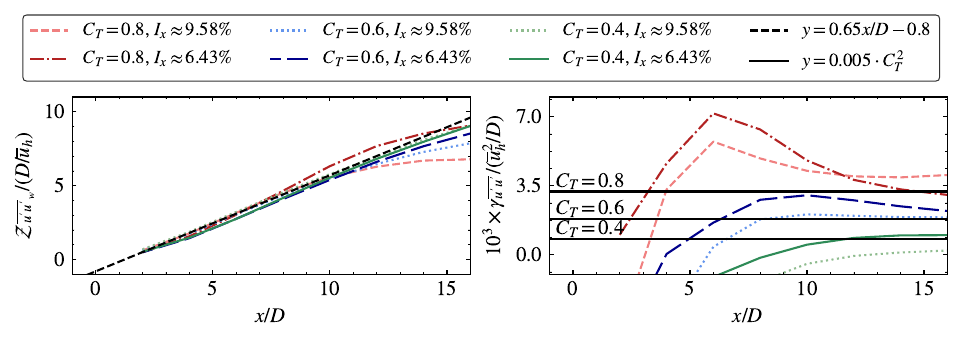}

    \caption{\tr{Streamwise evolution of $\mathcal{Z}_{\up{},w}$ (left) and $\gamma_{\up{}}$ (right) for different thrust coefficients and inflow turbulence intensities, shown alongside the proposed linear fit.}}
    \label{fig:zetaEvolution_upup}
\end{figure}

\tr{
The same procedure as for the~\gls{tke} analytical model is applied to calculate $\up{}_w$. The eddy viscosity in the inflow is computed using equation~\ref{eq:infiniteEddyViscosity}, with $l_{m,\infty}^{\text{max}} = 0.35\delta$. Figure~\ref{fig:zetaEvolution_upup} (left) shows the evolution of $\mathcal{Z}_{\up{},w}$ with respect to the streamwise distance. Once again, $\mathcal{Z}_{\up{},w}$ appears to be largely independent of the thrust coefficient and turbulence intensity, and is represented by a linear regression of the form $\mathcal{Z}_{\up{},w} / (D/u_h) = 0.65 x/D-0.8$. 
}

\tr{The evolution of $\gamma_{\up{}}$ is illustrated in figure~\ref{fig:zetaEvolution_upup} (right).}
\tr{
As for the turbulent kinetic energy, it highlights a strong dependence on the thrust coefficient and a weaker dependence on the turbulence intensity. Based on a similar reasoning as in section~\ref{subsec:zeta}, we model it as $\tr{\gamma_{\up{}} / (\bu_h^2/D) = 5 \times 10^{-3}\, C_T^2}$.}

\subsection{Comparison to \texorpdfstring{\gls{les}}{LES} data}
%

\begin{figure}[!ht]
    \centering
    \includegraphics[width=14cm]{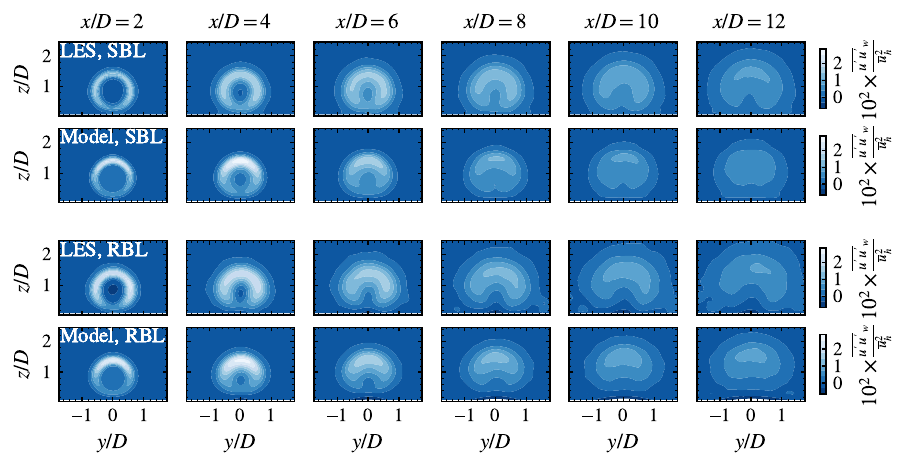}

    \caption{\tr{Contour plot of the wake-added streamwise normal turbulent shear stress: constant $x$ planes at several streamwise distances. SBL case (first row: \gls{les}, second row: analytical model) and the RBL case (third row: \gls{les}, fourth row: analytical model) from the calibration dataset.}}
    \label{fig:upupAnalyticalModelContours}
\end{figure}

\tr{We compare the analytical model predictions with the \gls{les} dataset for both the smooth and the rough boundary layers in figure~\ref{fig:upupAnalyticalModelContours}.} 
\tr{We note that, similarly to the previous section, the LES data was used to estimate the velocity field, which then served as input for the $\up{}_w$ field. 
Generally, we observe a qualitatively good agreement between the \gls{les} and the model predictions. }
\tr{Further validation of the model against experimental data will be performed in the next section.}





\section{Validations and comparison to two wind-tunnel datasets}\label{sec:windTunnel}
\tr{The analytical model for the normal turbulent stress is validated against data from two wind-tunnel campaigns covering inflow turbulence intensities from $I_x\approx 6.7\%$ to
$I_x\approx 13.8\%$.} 
\begin{table}
  \centering
  \renewcommand{\arraystretch}{1.8} 
  \caption{Description of the analytical models used in section \ref{sec:windTunnel} to predict the wake velocity and the wake-added turbulence.}
  \label{tab:unified_wake_model}
  \begin{tabular}{@{} l p{6.5cm} p{4.5cm} @{}}
    \toprule
    \textbf{Model Component} & \textbf{Equation / Definition} & \textbf{Constants \& Sub-parameters} \\
    \midrule
    
    \multicolumn{3}{@{}l}{\textit{\textbf{1. Wake Velocity Model}}} \\
    \midrule
    Streamwise  velocity $\bu{}$ & $\bu{}_\infty(z)\left(1- C(x)\exp\left(-\dfrac{1}{2} \dfrac{r^{n}}{\sigma^2}\right)\right)$ & $r=\sqrt{(y-y_{h})^2 + (z-z_h)^2}$ \\
    Inflow Profile $\bu{}_\infty$ & $\dfrac{\kappa}{u_*}\log\left(\dfrac{z}{z_0}\right)$ & \\
    Super-Gaussian Order $n$ & $a_f e^{b_f x}+c_f$ & $a_f = -8.2635 C_T^3 + 8.5939C_T^2$ \newline $\phantom{a_f =} - 8.9691C_T + 10.7286$ \newline $b_f = 1.68\exp(-25.98 I_x)-1.06$ \newline $c_f = 2$ \\    
    Wake Width ${\sigma}$ & $\left(a_s I_x+b_s\right)x+c_s\sqrt{\dfrac{1}{2}\dfrac{1+\sqrt{1-C_T}}{\sqrt{1-C_T}}}$ & $a_s = 0.28$ \newline $b_s = 0.01$ \newline $c_s = 0.1 C_T + 0.1$ \\
    Maximum velocity deficit $C(x)$ & $2^{2/n-1}-\sqrt{2^{4/n-2}-\dfrac{nC_T}{16\Gamma(2/n)\sigma^{4/n}}}$ & \\

    \midrule
    \multicolumn{3}{@{}l}{\textit{\textbf{2. Wake-Added Turbulence Model}}} \\
    \midrule
    Wake added $\up{}_w$ & $\zeta_{\up{},w} \left( \nu_{t,\infty} \left[\left(\dfrac{\partial \bu}{\partial y}\right)^2 +\left(\dfrac{\partial \bu}{\partial z}\right)^2 - \left(\dfrac{\partial \bu_\infty}{\partial z}\right)^2\right] - \gamma_{\up{}}\Delta \bu{} \right)$ & \\
    Parameter $\gamma_{\up{}}$ & $5 \times 10^{-3}\, C_T^2 (\bu_h^2/D)$ & \\
    Parameter $\zeta_{\up{},w}$ & $\begin{cases} \dfrac{z}{z_{BT}} (a_\zeta x +b_\zeta) & \text{if } 0<z\leq z_{BT} \\ a_\zeta x +b_\zeta & \text{if } z>z_{BT} \end{cases}$ & $a_\zeta=0.65$ \newline $b_\zeta=-0.8$ \\
    Inflow Eddy Viscosity $\nu_{t,\infty}$ & $\left(\dfrac{\kappa z}{1+\frac{\kappa z}{l_{m,\infty}^\text{max}}}\right)^2\dfrac{\partial \bu_\infty}{\partial z}$ & $l_{m,\infty}^\text{max}=0.35\delta$ \newline $\delta=3D$ (default) \\
    \bottomrule
  \end{tabular}
\end{table}
\tr{ Note that the model predictions rely on parameters calibrated against the LES data and assumed to be universal. Consequently, the wind-tunnel datasets were strictly ``unseen'' during model calibration and serve solely for validation purposes.}
Here, to have a complete engineering modelling framework and predict velocity distributions, we use the super-Gaussian wake model~\citep{Blondel2020}, although other wake velocity models \citep[e.g.,][among others]{Bastankhah2014,schreiber2020brief,Bastankhah2021, ali_diffusion-based_2024} could also be used. \tr{The parameters used in the present comparison are taken from \citet{Blondel2023}.}
\tr{Table~\ref{tab:unified_wake_model} provides a summary of the velocity deficit and wake-added turbulence models, together with all employed model parameters. The tilde superscript indicates normalization with respect to the turbine diameter $D$, and $\Gamma(\cdot)$ is the gamma function.}

\subsection{\tr{Wind tunnel experiments by \citet{Bastankhah2017}}}
\tr{The first test case considers the wind-tunnel data reported
in~\citet{Bastankhah2017}. A $\SI{0.15}{\metre}$ diameter miniature
wind turbine is immersed in a turbulent boundary layer with a hub-height turbulent
intensity of approximately $6.7\%$. The hub centre is located at $z_h=\SI{0.125}{\metre}$. In this case, the turbine operates at a high thrust
coefficient of $C_T\approx 0.77$.}

\begin{figure}[!ht]
        \centering
        \includegraphics[width=12cm]{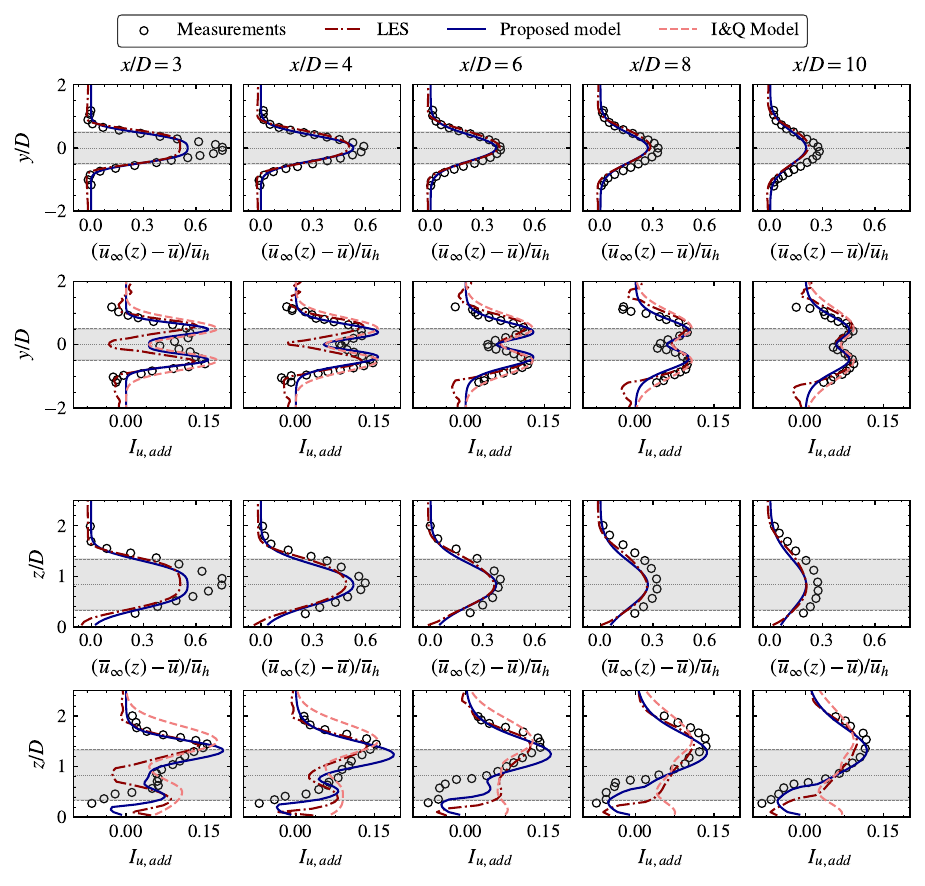}

    \caption{\tr{Lateral and vertical profiles of the velocity deficit (first and third rows, respectively) and wake-added turbulence intensity $I_{u,add}$ (second and fourth row, respectively): wind tunnel data from \citet{Bastankhah2017} and \up{} analytical model fed with super-Gaussian wake velocity model.}}    
    \label{fig:windTunnelBastankhahLateral}
\end{figure}

\tr{Figure~\ref{fig:windTunnelBastankhahLateral} 
shows the lateral and vertical profiles of normalised velocity deficit and wake-added turbulence intensity, respectively. The figures contain laboratory measurements, the LES data based on our LBM solver, and our proposed analytical model. They also include the added turbulence intensity predictions made by \cite{Ishihara2018}, shown as the I\&Q model.}
\tr{The figures demonstrate that the super-Gaussian model reproduces the velocity deficit in both lateral and vertical directions with good accuracy, enabling its application within the wake-added turbulence model.}

\tr{In terms of lateral wake-added turbulence intensity profiles, defined as $I_{u,add}=\sqrt{(\sigma_u/\bu)^2-(\sigma_{u_\infty}/\bu_\infty)^2}$ for $\sigma_u/\bu \ge \sigma_{u_\infty} / \bu_\infty$ and
$I_{u,add}=-\sqrt{(\sigma_{u_\infty}/\bu_\infty)^2-(\sigma_u/\bu)^2}$ otherwise, the model follows the experimental trends fairly closely over the entire set of streamwise distances. The agreement between the LES simulations and the analytical model is also very good, both following the experimental trends. In the near wake, at $x/D=3$ and $x/D=4$, an overestimation of the wake-added turbulence is observed in the~\gls{les}. This could be explained by the absence of a nacelle in the simplified actuator disk approach employed in the simulations.}
\tr{Below hub height, the I\&Q model overpredicts the wake-added turbulence intensity, for all downstream distances. As shown in the vertical profiles, 
there is a mismatch between the measurements and the~\gls{les} simulations close to the ground. Such a deviation was also observed in~\citet{Dar2024}. 
The overall behaviour of the proposed model for this first test case is highly satisfactory.}

\subsection{\tr{Wind tunnel experiments by \cite{Stein2019}}}
\tr{The second dataset, based on the experimental campaign reported in~\citep{Stein2019}}
\begin{figure}[!ht]
        \centering
        \includegraphics[width=13.5cm]{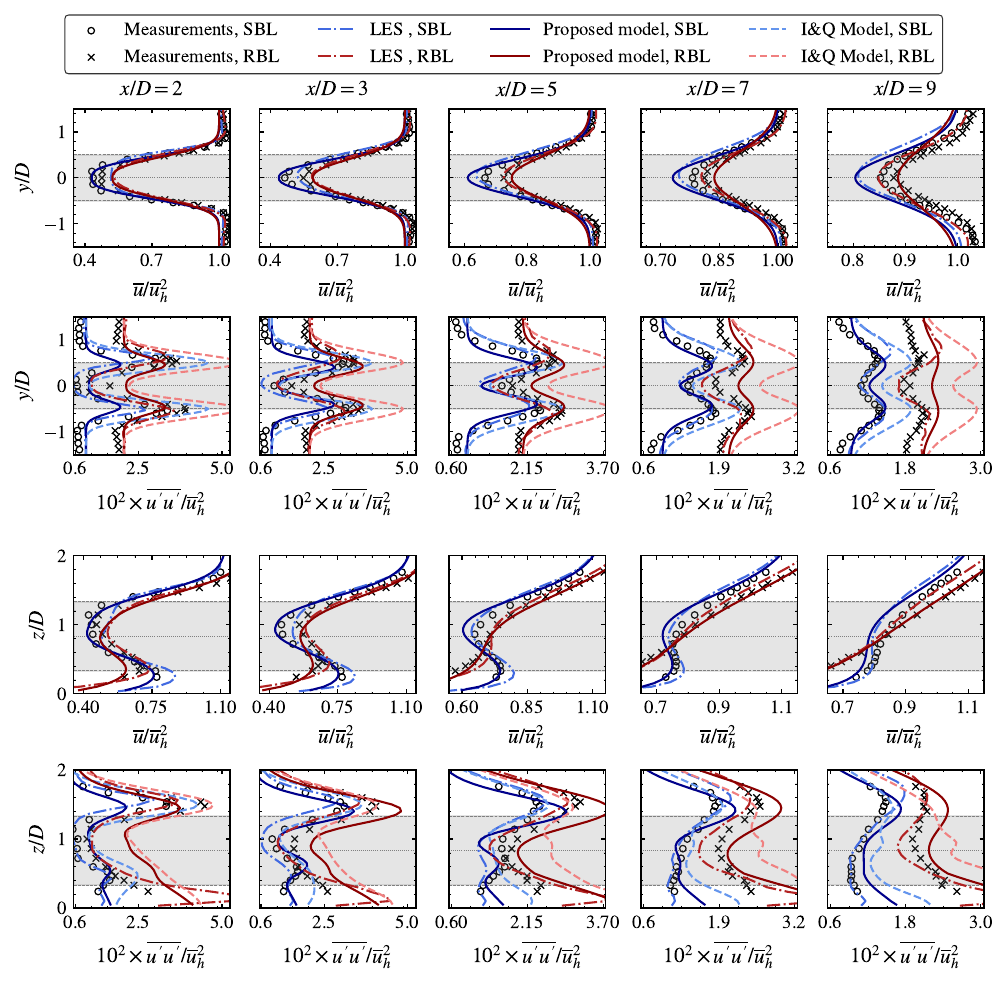}       
    \caption{\tr{Lateral and vertical profiles of the velocity deficit (first and third rows, respectively) and axial normal Reynolds stress $\up{}/\bu{}_h$ (second and fourth row, respectively): wind tunnel data from \citet{Stein2019} and \up{} analytical model fed with super-Gaussian wake velocity model.}}  
    
    \label{fig:windTunnelStein}
\end{figure}
\tr{, allows for a direct assessment of the influence of the inflow turbulence intensity. Two inflow conditions are
examined: a \gls{sbl} with $TI\approx 8.3\%$ and a \gls{rbl} with $TI\approx 13.8\%$. The rotor diameter and hub centre are equal, i.e., $z_h=D=\SI{0.45}{\metre}$.}

\tr{As shown in figure~\ref{fig:windTunnelStein}, the super-Gaussian model shows good agreement with the reference data and provides reliable input for the wake-added turbulence model.
In the lateral direction, figure~\ref{fig:windTunnelStein}, the proposed model is able to correctly reproduce the wake-added streamwise normal Reynolds stress behaviour for both~\gls{abl}s, highlighting its robustness. The I\&Q model reproduces the~\gls{sbl} case very accurately, but strongly overestimates the streamwise normal Reynolds stress in the~\gls{rbl} case.}
\tr{In the vertical direction, figure~\ref{fig:windTunnelStein}, similar observations are drawn. In the near-wake and below the top-tip region, a clear overestimation of the wake-added $\up{}$ is observed in the~\gls{rbl} case. In the rest of the wake, the proposed approach slightly overestimates the turbulence at the top tip position, but accurately reproduces the rest of the profile. The I\&Q model underpredicts the wake-added streamwise normal Reynolds stress below the top-tip region in the far wake, $x/D\ge 5$, in the~\gls{rbl} case, but accurately reproduces the~\gls{sbl} case. The~\gls{les} simulation is very close to the measured data, even in the near-wake region. }


\section{Conclusion and future work}
In this work, we derived an analytical model to predict the \gls{tke} and the normal turbulent stress and validated the results against \acrlongpl{les} and wind tunnel measurements. It is grounded in the analysis of turbulence transport equation budgets, classical approximations commonly used in eddy-viscosity \gls{rans} closures, and additional simplifications leading to a tractable analytical formulation. 
The derivation achieved several intermediate results. Firstly, we formulated a simplified transport equation for wake-added \gls{tke}, which supports initial assumptions and can be used directly due to its low computational cost.
From this equation, we then use far-wake assumptions to obtain an analytical model for wake-added \gls{tke} by applying a self-similarity hypothesis for the advection term and a budget-based simplification of the diffusion term. 
We then applied the same methodology to the $\up{}_w$ transport equation, resulting in a final model structurally similar to the \gls{tke} model.
This model can replace empirical turbulence models in analytical wind farm solvers. Its reliance on physical principles and its ability to incorporate inflow shear make it a powerful predictive tool. 
\tr{Future work will focus on extending the model to farm-scale wake predictions and the inclusion of atmospheric stability. 
\revTwo{Finally, it is worth noting that the wake recovery rate is expected to depend on the local level of turbulence rather than the inflow value~\citep{Pedersen2022}. Accounting for this can be achieved in future work by coupling a wake velocity model with the turbulence model developed in this study via an iterative approach.}
}

\appendix
\section{\tr{Impact of the Reynolds number on the~\gls{les} simulations}}\label{app:lesValidation}

\tr{In the setup used throughout this work, wind-tunnel-scale simulations are performed, based on the solver validation against the experimental data from~\citet{Bastankhah2017}. Here, an additional simulation at utility scale is presented, using a rotor diameter of $D=\SI{150}{\meter}$ and a velocity at hub-height of $\SI{8}{\meter\per\second}$. The other parameters are kept constant. With this setup, we want to assess the impact of the rotor diameter (more generally, of the Reynolds number) on our simulations. At the wind tunnel scale, the Reynolds number based on the rotor diameter is of roughly $2\times 10^4$, while at the utility scale, it reaches $8\times 10^7$.}

\begin{figure}[!ht]
    \centering
    \includegraphics[width=13.7cm]{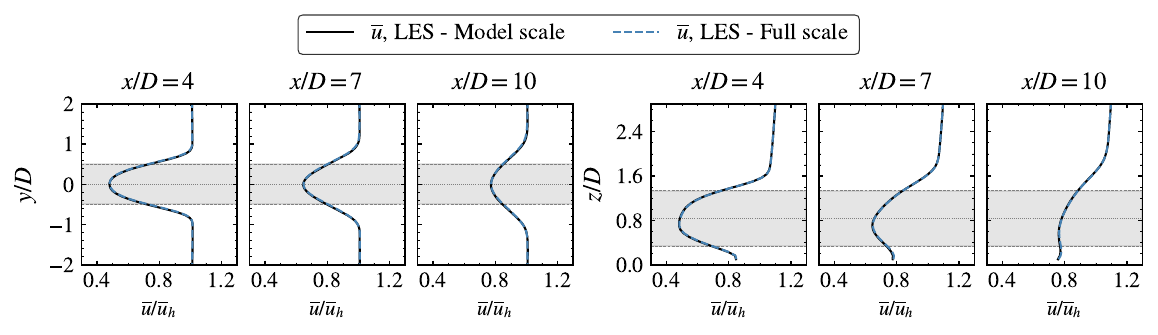}
    \caption{\tr{Lateral and vertical velocity profiles at full scale and model scale.}}
    \label{fig:scaleImpact_U}
\end{figure}

\begin{figure}[!ht]
    \centering
    \includegraphics[width=13.7cm]{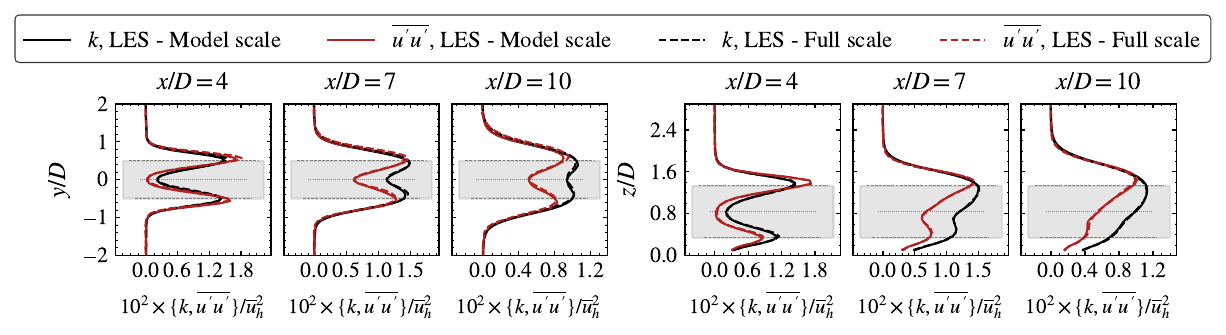}
    \caption{\tr{Lateral and vertical~\gls{tke} and streamwise normal Reynolds stress profiles at full scale and model scale.}}
    \label{fig:scaleImpact_turbulence}
\end{figure}

\tr{Figure~\ref{fig:scaleImpact_U} shows that the impact of the Reynolds number on the predicted velocity is largely negligible: the profiles are almost perfectly superimposed.}
\tr{Similarly, figure~\ref{fig:scaleImpact_turbulence} shows the impact of the Reynolds number on the~\gls{tke} and on the streamwise component of the normal Reynolds stress. The resulting profiles are essentially independent of the Reynolds number. In an experimental study with a scaled wind turbine, \citet{Chamorro2012} observed a good level of Reynolds-number independence only for slightly higher Reynolds numbers. However, as noted in \citep{Chamorro2012}, the wake in the wind tunnel might be influenced by the Reynolds-number-dependent local turbine airfoil aerodynamics, which does not occur in our actuator-disk simulations.}

\section{Verification of the advection, diffusion and production terms treatment}\label{app:APT}

\begin{figure}[!ht]
    \centering
    \includegraphics[width=14cm]{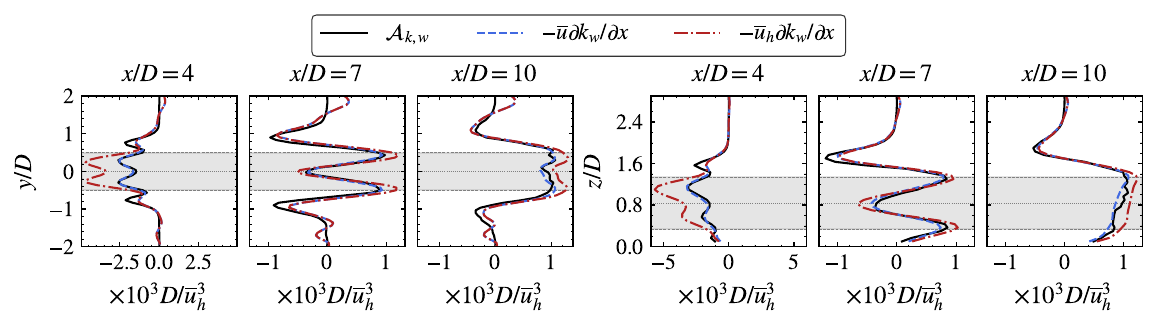}
    \caption{Lateral and vertical profiles of the \gls{tke} advection term: comparison between the complete and the simplified terms based on \gls{les} data.}
    \label{fig:simpleAdvection}
\end{figure}

This appendix assesses the validity of some of the assumptions used to simplify the \gls{tke} transport equations in section \ref{sec:term-by_term}. Figure~\ref{fig:simpleAdvection} shows the full advection term\tr{, }the simplified advection term\tr{,  which consists only of $\bu \partial k_w /\partial x$, and the linearised term, $\bu_h \partial k_w /\partial x$}. Despite some numerical oscillations at the outer edges of the wake, the overall conclusion remains unaffected: using only one term is sufficient to model the advection accurately. \tr{Furthermore, linearising the advection term by replacing $\bu$ with $\bu_h$ has only a marginal impact in the far-wake.}

\begin{figure}[!ht]
    \centering
    \includegraphics[width=14cm]{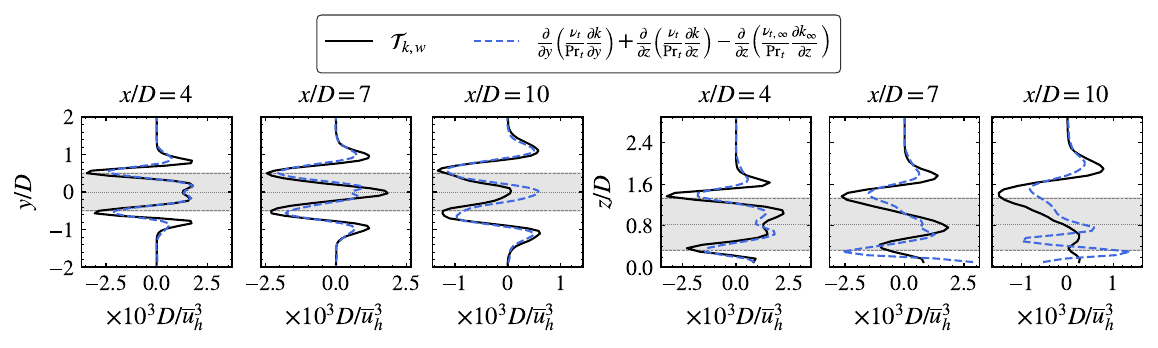}
    \caption{Lateral and vertical profiles of the \gls{tke} diffusion term: comparison between the complete and the simplified terms based on \gls{les} data.}
    \label{fig:simpleDiffusion}
\end{figure}

Figure~\ref{fig:simpleDiffusion} compares the proposed model for the turbulent transport term and the complete term. Overall, the agreement between the chosen transport model and the complete term is less good than for the advection and production term. This discrepancy could originate, at least partially, from numerical errors introduced when computing the second-order derivative of the wake-added \gls{tke}.

The last term discussed in the appendix is the production term. Figure~\ref{fig:simpleProduction} compares the complete production term to its model based on the Boussinesq eddy viscosity hypothesis and a simplified rate-of-strain tensor. The eddy viscosities in the model is obtained directly from the \gls{les} data as $\nu_t = \mathcal{P}_k / (2S_{ij}S_{ij})$\tr{.}
\begin{figure}[!ht]
    \centering
    \includegraphics[width=14cm]{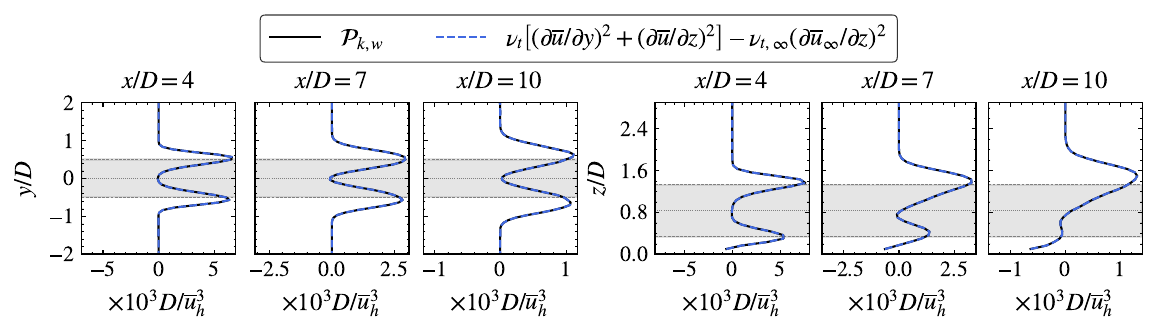}
    \caption{Lateral and vertical profiles of the \gls{tke} production term: comparison between the complete and the simplified terms based on \gls{les} data.}
    \label{fig:simpleProduction}
\end{figure}
\tr{The model reproduces the \gls{les} data very well.} 

\section{\tr{Additional results: turbulent time scale evolution}}\label{app:TauEvolution}

\tr{In section~\ref{sec:dissipation-time-scale-models}, a simple model for the turbulent time scale $\tau_k$ in a wind turbine wake is introduced, based on the assumption that $\tau_k$ remains constant within the wake. The validity of this assumption is further assessed here using additional simulation cases.}

\begin{figure}[!ht]
    \centering
    \includegraphics[width=13.7cm]{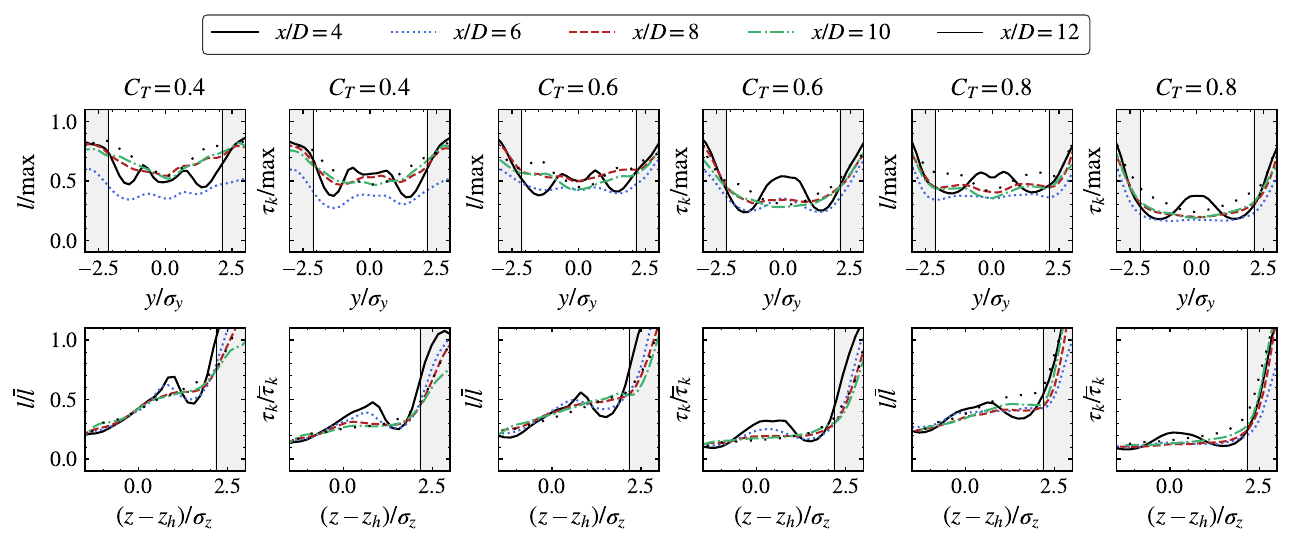}    
    \caption{\tr{Lateral (top) and vertical (bottom) self-similar profiles of the turbulent length scale $l$ and time scale $\tau_k$: SBL case. The shaded areas indicate the regions outside of the wake, where the normalised velocity deficit is below one percent.}}
    \label{fig:selfSimilarTau_SBL}
\end{figure}

\begin{figure}[!ht]
    \centering
    \includegraphics[width=13.7cm]{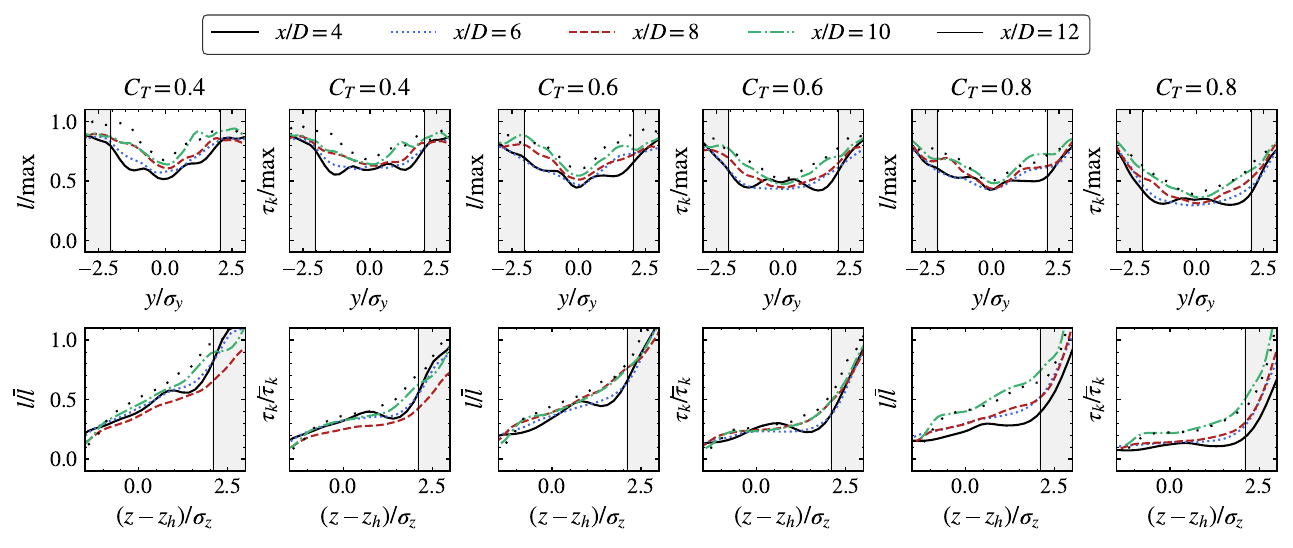}    
    \caption{\tr{Lateral (top) and vertical (bottom) self-similar profiles of the turbulent length scale $l$ and time scale $\tau_k$: RBL case. The shaded areas indicate the regions outside of the wake, where the normalised velocity deficit is below one per cent.}}
    \label{fig:selfSimilarTau_RBL}
\end{figure}

\tr{Figures~\ref{fig:selfSimilarTau_SBL} and~\ref{fig:selfSimilarTau_RBL} present the evolution of $\tau_k$ for three different thrust coefficients $C_T$, plotted as a function of the lateral and vertical distances normalised by the Gaussian wake width $\sigma$. The evolution of the turbulent length scale $l = k^{3/2}/\varepsilon$ is also shown. Since $l$ is often assumed to be constant in wake modelling (see, e.g.,~\citet{Klemmer2025,Bastankhah2024}), comparing its behaviour with that of $\tau_k$ is of particular interest. Figure~\ref{fig:selfSimilarTau_SBL} corresponds to the \gls{sbl} case, while figure~\ref{fig:selfSimilarTau_RBL} focuses on the \gls{rbl} case.}

\tr{For the lateral profiles, all quantities are normalised by their respective maximum values, as is customary when analysing self-similar behaviour. For the vertical profiles, normalisation is performed using the mean value over the profile. This choice avoids excessive compression of the curves and improves the readability of the figures.}
\tr{Both figures indicate that treating $\tau_k$ as constant within the wake region is a reasonable approximation, regardless of ground roughness. Moreover, the validity of this assumption appears to improve with increasing thrust coefficient.}

\tr{The same figures also suggest that assuming a constant turbulent length scale $l$ within the wake is more questionable. While this approximation seems acceptable in the lateral direction, significant variations of $l$ are observed in the vertical direction, with a noticeable increase with height.}

\clearpage

\bibliographystyle{jfm}
\bibliography{jfm}

\begin{thebibliography}{76}
\expandafter\ifx\csname natexlab\endcsname\relax\def\natexlab#1{#1}\fi
\def\au#1{#1} \def\ed#1{#1} \def\yr#1{#1}\def\at#1{#1}\def\jt#1{\textit{#1}}
  \def\bt#1{#1}\def\bvol#1{\textbf{#1}} \def\vol#1{#1} \def\pg#1{#1}
  \def\publ#1{#1}\def\arxiv#1{#1}\def\org#1{#1}\def\st#1{\textit{#1}}

\bibitem[Ali {\em et~al.\/}(2024)Ali, Stallard \&
  Ouro]{ali_diffusion-based_2024}
{\sc \au{Ali, Karim}, \au{Stallard, Tim} \& \au{Ouro, Pablo}} \yr{2024}  \at{A
  diffusion-based wind turbine wake model}.  \jt{Journal of Fluid Mechanics}
  \bvol{1001},  \pg{A13}.

\bibitem[Apsley \& Castro(1997)]{Apsley1997}
{\sc \au{Apsley, David} \& \au{Castro, {I P}}} \yr{1997}  \at{A
  limited-length-scale k- model for the neutral and stably-stratified
  atmospheric boundary layer,}.  \jt{Boundary-Layer Meteorology}
  \bvol{83}~(0),  \pg{75--98}.

\bibitem[Aubrun {\em et~al.\/}(2013)Aubrun, Loyer, Hancock \&
  Hayden]{Aubrun2013}
{\sc \au{Aubrun, Sandrine}, \au{Loyer, St\'ephane}, \au{Hancock, Philip~E.} \&
  \au{Hayden, Paul}} \yr{2013}  \at{Wind turbine wake properties: Comparison
  between a non-rotating simplified wind turbine model and a rotating model}.
  \jt{Journal of Wind Engineering and Industrial Aerodynamics}  \bvol{120},
  \pg{1--8}.

\bibitem[Bastankhah \& Port{\'e}-Agel(2017)]{bastankhah2017new}
{\sc \au{Bastankhah, Majid} \& \au{Port{\'e}-Agel, Fernando}} \yr{2017}  \at{A
  new miniature wind turbine for wind tunnel experiments. part ii: wake
  structure and flow dynamics}.  \jt{Energies}  \bvol{10}~(7),  \pg{923}.

\bibitem[Bastankhah \& Porté-Agel(2014)]{Bastankhah2014}
{\sc \au{Bastankhah, Majid} \& \au{Porté-Agel, Fernando}} \yr{2014}  \at{A new
  analytical model for wind-turbine wakes}.  \jt{Renewable Energy}  \bvol{70},
  \pg{116--123}, special issue on aerodynamics of offshore wind energy systems
  and wakes.

\bibitem[Bastankhah \& Porté-Agel(2016)]{Bastankhah2016}
{\sc \au{Bastankhah, Majid} \& \au{Porté-Agel, Fernando}} \yr{2016}
  \at{Experimental and theoretical study of wind-turbine wakes in yawed
  conditions}.  \jt{Journal of Fluid Mechanics}  \bvol{806},  \pg{506–541}.

\bibitem[Bastankhah \& Porté-Agel(2017)]{Bastankhah2017}
{\sc \au{Bastankhah, Majid} \& \au{Porté-Agel, Fernando}} \yr{2017}  \at{A new
  miniature wind turbine for wind tunnel experiments. part ii: Wake structure
  and flow dynamics}.  \jt{Energies}  \bvol{10}~(7).

\bibitem[Bastankhah {\em et~al.\/}(2021)Bastankhah, Welch, Martínez-Tossas,
  King \& Fleming]{Bastankhah2021}
{\sc \au{Bastankhah, Majid}, \au{Welch, Bridget~L.}, \au{Martínez-Tossas,
  Luis~A.}, \au{King, Jennifer} \& \au{Fleming, Paul}} \yr{2021}
  \at{Analytical solution for the cumulative wake of wind turbines in wind
  farms}.  \jt{Journal of Fluid Mechanics}  \bvol{911},  \pg{A53}.

\bibitem[Bastankhah {\em et~al.\/}(2024)Bastankhah, Zunder, Hydon, Deebank \&
  Placidi]{Bastankhah2024}
{\sc \au{Bastankhah, Majid}, \au{Zunder, Jenna~K.}, \au{Hydon, Peter~E.},
  \au{Deebank, Charles} \& \au{Placidi, Marco}} \yr{2024}  \at{Modelling
  turbulence in axisymmetric wakes: an application to wind turbine wakes}.
  \jt{Journal of Fluid Mechanics}  \bvol{1000},  \pg{A2}.

\bibitem[Bauer {\em et~al.\/}(2021)Bauer, Eibl, Godenschwager, Kohl, Kuron,
  Rettinger, Schornbaum, Schwarzmeier, Thönnes, Köstler \& Rüde]{Bauer2021}
{\sc \au{Bauer, Martin}, \au{Eibl, Sebastian}, \au{Godenschwager, Christian},
  \au{Kohl, Nils}, \au{Kuron, Michael}, \au{Rettinger, Christoph},
  \au{Schornbaum, Florian}, \au{Schwarzmeier, Christoph}, \au{Thönnes,
  Dominik}, \au{Köstler, Harald} \& \au{Rüde, Ulrich}} \yr{2021}
  \at{walberla: A block-structured high-performance framework for multiphysics
  simulations}.  \jt{Computers \& Mathematics with Applications}  \bvol{81},
  \pg{478--501}, development and Application of Open-source Software for
  Problems with Numerical PDEs.

\bibitem[Bauer \& Rüde(2018)]{Bauer2018}
{\sc \au{Bauer, Martin} \& \au{Rüde, Ulrich}} \yr{2018} An improved lattice
  boltzmann d3q19 method based on an alternative equilibrium discretization,
  \arxiv{arXiv: 1803.04937}.

\bibitem[Bay {\em et~al.\/}(2023)Bay, Fleming, Doekemeijer, King, Churchfield
  \& Mudafort]{Bay2023}
{\sc \au{Bay, Christopher.~J.}, \au{Fleming, Paul}, \au{Doekemeijer, Bart},
  \au{King, Jennifer}, \au{Churchfield, Matt} \& \au{Mudafort, Rafael}}
  \yr{2023}  \at{Addressing deep array effects and impacts to wake steering
  with the cumulative-curl wake model}.  \jt{Wind Energy Science}
  \bvol{8}~(3),  \pg{401--419}.

\bibitem[Blackadar(1962)]{Blackadar1962}
{\sc \au{Blackadar, Alfred~K.}} \yr{1962}  \at{The vertical distribution of
  wind and turbulent exchange in a neutral atmosphere}.  \jt{Journal of
  Geophysical Research (1896-1977)}  \bvol{67}~(8),  \pg{3095--3102},
  \arxiv{arXiv:
  https://agupubs.onlinelibrary.wiley.com/doi/pdf/10.1029/JZ067i008p03095}.

\bibitem[Blondel(2023)]{Blondel2023}
{\sc \au{Blondel, Fr\'ed\'eric}} \yr{2023}  \at{Brief communication: A
  momentum-conserving superposition method applied to the super-gaussian wind
  turbine wake model}.  \jt{Wind Energy Science}  \bvol{8}~(2),  \pg{141--147}.

\bibitem[Blondel \& Cathelain(2020)]{Blondel2020}
{\sc \au{Blondel, Frédéric} \& \au{Cathelain, Marie}} \yr{2020}  \at{An
  alternative form of the super-gaussian wind turbine wake model}.  \jt{Wind
  Energy Science}  \bvol{5}~(3),  \pg{1225--1236}.

\bibitem[Chamorro {\em et~al.\/}(2012)Chamorro, Arndt \&
  Sotiropoulos]{Chamorro2012}
{\sc \au{Chamorro, Leonardo.}, \au{Arndt, R.E.A} \& \au{Sotiropoulos, F.}}
  \yr{2012}  \at{Reynolds number dependence of turbulence statistics in the
  wake of wind turbines}.  \jt{Wind Energy}  \bvol{15}~(5),  \pg{733--742},
  \arxiv{arXiv: https://onlinelibrary.wiley.com/doi/pdf/10.1002/we.501}.

\bibitem[Chamorro \& Porté-Agel(2008)]{Chamorro2008}
{\sc \au{Chamorro, Leonardo} \& \au{Porté-Agel, Fernando}} \yr{2008}  \at{A
  wind-tunnel investigation of wind-turbine wakes: Boundary-layer turbulence
  effects}.  \jt{Boundary-Layer Meteorology}  \bvol{132},  \pg{129--149}.

\bibitem[Colombi\'e {\em et~al.\/}(2021)Colombi\'e, Laroche, Chedevergne,
  Manceau, Duchaine \& Gicquel]{Colombie2021}
{\sc \au{Colombi\'e, A.}, \au{Laroche, E.}, \au{Chedevergne, F.}, \au{Manceau,
  R.}, \au{Duchaine, F.} \& \au{Gicquel, L.}} \yr{2021}
  \at{Large-eddy-simulation-based analysis of reynolds-stress budgets for a
  round impinging jet}.  \jt{Physics of Fluids}  \bvol{33}~(11),  \pg{115109}.

\bibitem[Crespo \& Hernández(1996)]{Crespo1996}
{\sc \au{Crespo, Antonio} \& \au{Hernández, Julio}} \yr{1996}  \at{Turbulence
  characteristics in wind-turbine wakes}.  \jt{Journal of Wind Engineering and
  Industrial Aerodynamics}  \bvol{61}~(1),  \pg{71--85}.

\bibitem[Dar {\em et~al.\/}(2024)Dar, Majzoub \& Porté-Agel]{Dar2024}
{\sc \au{Dar, Arslan~Salim}, \au{Majzoub, Rim} \& \au{Porté-Agel, Fernando}}
  \yr{2024}  \at{The effect of nacelle-to-rotor size on the wake of a miniature
  wind turbine}.  \jt{Journal of Physics: Conference Series}  \bvol{2767}~(9),
  \pg{092057}.

\bibitem[Dhamankar {\em et~al.\/}(2018)Dhamankar, Blaisdell \&
  Lyrintzis]{Dhamankar2018}
{\sc \au{Dhamankar, Nitin~S.}, \au{Blaisdell, Gregory~A.} \& \au{Lyrintzis,
  Anastasios~S.}} \yr{2018}  \at{Overview of turbulent inflow boundary
  conditions for large-eddy simulations}.  \jt{AIAA Journal}  \bvol{56}~(4),
  \pg{1317--1334},  \arxiv{arXiv: https://doi.org/10.2514/1.J055528}.

\bibitem[Frandsen(1992)]{Frandsen1992}
{\sc \au{Frandsen, Sten}} \yr{1992}  \at{On the wind speed reduction in the
  center of large clusters of wind turbines}.  \jt{Journal of Wind Engineering
  and Industrial Aerodynamics}  \bvol{39}~(1),  \pg{251--265}.

\bibitem[Frandsen {\em et~al.\/}(2006)Frandsen, Barthelmie, Pryor, Rathmann,
  Larsen, H{\o}jstrup \& Th{\o}gersen]{Frandsen2006}
{\sc \au{Frandsen, Sten}, \au{Barthelmie, Rebecca}, \au{Pryor, Sara},
  \au{Rathmann, Ole}, \au{Larsen, S{\o}ren}, \au{H{\o}jstrup, J{\o}rgen} \&
  \au{Th{\o}gersen, Morten}} \yr{2006}  \at{Analytical modelling of wind speed
  deficit in large offshore wind farms}.  \jt{Wind Energy: An International
  Journal for Progress and Applications in Wind Power Conversion Technology}
  \bvol{9}~(1-2),  \pg{39--53}.

\bibitem[Gehrke \& Rung(2022)]{Gehrke2022}
{\sc \au{Gehrke, Martin} \& \au{Rung, Thomas}} \yr{2022}  \at{Scale-resolving
  turbulent channel flow simulations using a dynamic cumulant lattice boltzmann
  method}.  \jt{Physics of Fluids}  \bvol{34}~(7),  \pg{075129}.

\bibitem[Geier {\em et~al.\/}(2017)Geier, Pasquali \& Schönherr]{Geier2017}
{\sc \au{Geier, Martin}, \au{Pasquali, Andrea} \& \au{Schönherr, Martin}}
  \yr{2017}  \at{Parametrization of the cumulant lattice boltzmann method for
  fourth order accurate diffusion part i: Derivation and validation}.
  \jt{Journal of Computational Physics}  \bvol{348},  \pg{862--888}.

\bibitem[Geier {\em et~al.\/}(2015)Geier, Schönherr, Pasquali \&
  Krafczyk]{Geier2015}
{\sc \au{Geier, Martin}, \au{Schönherr, Martin}, \au{Pasquali, Andrea} \&
  \au{Krafczyk, Manfred}} \yr{2015}  \at{The cumulant lattice boltzmann
  equation in three dimensions: Theory and validation}.  \jt{Computers \&
  Mathematics with Applications}  \bvol{70}~(4),  \pg{507--547}.

\bibitem[Gerolymos \& Vallet(2016)]{Gerolymos2016}
{\sc \au{Gerolymos, Georges~A.} \& \au{Vallet, Isabelle}} \yr{2016}  \at{The
  dissipation tensor $\epsilon_{ij}$ in wall turbulence}.  \jt{Journal of Fluid
  Mechanics}  \bvol{807},  \pg{386–418}.

\bibitem[Han {\em et~al.\/}(2021)Han, Ooka \& Kikumoto]{Han2021}
{\sc \au{Han, Mengtao}, \au{Ooka, Ryozo} \& \au{Kikumoto, Hideki}} \yr{2021}
  \at{A wall function approach in lattice boltzmann method: algorithm and
  validation using turbulent channel flow}.  \jt{Fluid Dynamics Research}
  \bvol{53}~(4),  \pg{045506}.

\bibitem[Hancock {\em et~al.\/}(2014)Hancock, Zhang, Pascheke \&
  Hayden]{Hancock2014}
{\sc \au{Hancock, P~E}, \au{Zhang, S}, \au{Pascheke, F} \& \au{Hayden, P}}
  \yr{2014}  \at{Wind tunnel simulation of a wind turbine wake in neutral,
  stable and unstable wind flow}.  \jt{Journal of Physics: Conference Series}
  \bvol{555}~(1),  \pg{012047}.

\bibitem[Hanjalić \& Launder(1976)]{Hanjalic1976}
{\sc \au{Hanjalić, Kemal} \& \au{Launder, Brian~E.}} \yr{1976}
  \at{Contribution towards a reynolds-stress closure for low-reynolds-number
  turbulence}.  \jt{Journal of Fluid Mechanics}  \bvol{74}~(4),
  \pg{593–610}.

\bibitem[Harlow \& Hirt(1969)]{Harlow1969}
{\sc \au{Harlow, Francis~H} \& \au{Hirt, Cyril~W}} \yr{1969} {\em Generalized
  transport theory of anisotropic turbulence\/}.  \publ{Springfield, Va. :
  National Technical Information Service}, at head of title: Los Alamos
  Scientific Laboratory of the University of California, Los Alamos, New
  Mexico. La-4086; Uc-34, Physics; Tid-4500.

\bibitem[Heinze {\em et~al.\/}(2015)Heinze, Mironov \& Raasch]{Heinze2015}
{\sc \au{Heinze, Rieke}, \au{Mironov, Dmitrii} \& \au{Raasch, Siegfried}}
  \yr{2015}  \at{Second-moment budgets in cloud topped boundary layers: A
  large-eddy simulation study}.  \jt{Journal of Advances in Modeling Earth
  Systems}  \bvol{7}~(2),  \pg{510--536},  \arxiv{arXiv:
  https://agupubs.onlinelibrary.wiley.com/doi/pdf/10.1002/2014MS000376}.

\bibitem[Ib \& Lundtang~Petersen(1989)]{EWA_1989}
{\sc \au{Ib, Troen} \& \au{Lundtang~Petersen, Erik}} \yr{1989} {\em European
  Wind Atlas\/}.  \publ{Ris{\o} National Laboratory}.

\bibitem[Ishihara \& Qian(2018)]{Ishihara2018}
{\sc \au{Ishihara, Takeshi} \& \au{Qian, Guo-Wei}} \yr{2018}  \at{A new
  gaussian-based analytical wake model for wind turbines considering ambient
  turbulence intensities and thrust coefficient effects}.  \jt{Journal of Wind
  Engineering and Industrial Aerodynamics}  \bvol{177},  \pg{275--292}.

\bibitem[Iungo {\em et~al.\/}(2017)Iungo, Santhanagopalan, Ciri, Viola, Zhan,
  Rotea \& Leonardi]{Iungo2017}
{\sc \au{Iungo, Giacomo}, \au{Santhanagopalan, Vignesh}, \au{Ciri, Umberto},
  \au{Viola, Francesco}, \au{Zhan, Lu}, \au{Rotea, Mario} \& \au{Leonardi,
  Stefano}} \yr{2017}  \at{Parabolic rans solver for low‐computational‐cost
  simulations of wind turbine wakes}.  \jt{Wind Energy}  \bvol{21}.

\bibitem[J\'ez\'equel {\em et~al.\/}(2024)J\'ez\'equel, Blondel \&
  Masson]{Jezequel2024b}
{\sc \au{J\'ez\'equel, Erwan}, \au{Blondel, Fr\'ed\'eric} \& \au{Masson,
  Valery}} \yr{2024}  \at{Breakdown of the velocity and turbulence in the wake
  of a wind turbine -- part 2: Analytical modeling}.  \jt{Wind Energy Science}
  \bvol{9}~(1),  \pg{119--139}.

\bibitem[Jones \& Launder(1972)]{Jones1972}
{\sc \au{Jones, William~P.} \& \au{Launder, Brian~E.}} \yr{1972}  \at{The
  prediction of laminarization with a two-equation model of turbulence}.
  \jt{International Journal of Heat and Mass Transfer}  \bvol{15}~(2),
  \pg{301--314}.

\bibitem[Katic {\em et~al.\/}(1987)Katic, H{\o}jstrup \& Jensen]{Katic1987}
{\sc \au{Katic, I.}, \au{H{\o}jstrup, J.} \& \au{Jensen, N.O.}} \yr{1987} A
  simple model for cluster efficiency.  \bt{In {\em EWEC'86. Proceedings. Vol.
  1\/} (ed. \ed{W.~Palz \& E.~Sesto})},  \pg{pp. 407--410}.  \publ{A. Raguzzi},
  european Wind Energy Association Conference and Exhibition, EWEC '86 ;
  Conference date: 06-10-1986 Through 08-10-1986.

\bibitem[Keane {\em et~al.\/}(2016)Keane, Aguirre, Ferchland, Clive \&
  Gallacher]{Keane2016}
{\sc \au{Keane, Aidan}, \au{Aguirre, Pablo E.~Olmos}, \au{Ferchland, Hannah},
  \au{Clive, Peter} \& \au{Gallacher, Daniel}} \yr{2016}  \at{An analytical
  model for a full wind turbine wake}.  \jt{Journal of Physics: Conference
  Series}  \bvol{753}~(3),  \pg{032039}.

\bibitem[Khanjari {\em et~al.\/}(2025)Khanjari, Feroz \& Archer]{Khanjari2025}
{\sc \au{Khanjari, A.}, \au{Feroz, A.} \& \au{Archer, C.~L.}} \yr{2025}  \at{An
  analytical formulation for turbulence kinetic energy added by wind turbines
  based on large-eddy simulation}.  \jt{Wind Energy Science}  \bvol{10}~(5),
  \pg{887--905}.

\bibitem[Klemmer \& Howland(2024)]{Klemmer2024}
{\sc \au{Klemmer, Kerry~S.} \& \au{Howland, Michael~F.}} \yr{2024}
  \at{Momentum deficit and wake-added turbulence kinetic energy budgets in the
  stratified atmospheric boundary layer}.  \jt{Phys. Rev. Fluids}  \bvol{9},
  \pg{114607}.

\bibitem[Klemmer \& Howland(2025)]{Klemmer2025}
{\sc \au{Klemmer, Kerry~S.} \& \au{Howland, Michael~F.}} \yr{2025}  \at{Wake
  turbulence modeling in stratified atmospheric flows using a novel k-l model}.
   \jt{Journal of Renewable and Sustainable Energy}  \bvol{17}~(3),
  \pg{033304},  \arxiv{arXiv:
  https://pubs.aip.org/aip/jrse/article-pdf/doi/10.1063/5.0249278/20551084/033304\_1\_5.0249278.pdf}.

\bibitem[van~der Laan {\em et~al.\/}(2015)van~der Laan, Sørensen, Réthoré,
  Mann, Kelly, Troldborg, Schepers \& Machefaux]{vanDerLaan2015}
{\sc \au{van~der Laan, M.~Paul}, \au{Sørensen, Niels~N.}, \au{Réthoré,
  Pierre-Elouan}, \au{Mann, Jakob}, \au{Kelly, Mark~C.}, \au{Troldborg, Niels},
  \au{Schepers, J.~Gerard} \& \au{Machefaux, Ewan}} \yr{2015}  \at{An improved
  k-$\varepsilon$ model applied to a wind turbine wake in atmospheric
  turbulence}.  \jt{Wind Energy}  \bvol{18}~(5),  \pg{889--907},  \arxiv{arXiv:
  https://onlinelibrary.wiley.com/doi/pdf/10.1002/we.1736}.

\bibitem[Launder \& Spalding(1974)]{Launder1974}
{\sc \au{Launder, B.E.} \& \au{Spalding, D.B.}} \yr{1974}  \at{The numerical
  computation of turbulent flows}.  \jt{Computer Methods in Applied Mechanics
  and Engineering}  \bvol{3}~(2),  \pg{269--289}.

\bibitem[Launder {\em et~al.\/}(1975)Launder, Reece \& Rodi]{Launder1975}
{\sc \au{Launder, B.~E.}, \au{Reece, G.~J.} \& \au{Rodi, W.}} \yr{1975}
  \at{Progress in the development of a reynolds-stress turbulence closure}.
  \jt{Journal of Fluid Mechanics}  \bvol{68}~(3),  \pg{537–566}.

\bibitem[Lin \& Porté-Agel(2019)]{MouLin2019}
{\sc \au{Lin, Mou} \& \au{Porté-Agel, Fernando}} \yr{2019}  \at{Large-eddy
  simulation of yawed wind-turbine wakes: Comparisons with wind tunnel
  measurements and analytical wake models}.  \jt{Energies}  \bvol{12}~(23).

\bibitem[Lumley(1975)]{Lumley1975}
{\sc \au{Lumley, John~L.}} \yr{1975}  \at{Pressure‐strain correlation}.
  \jt{The Physics of Fluids}  \bvol{18}~(6),  \pg{750--750},  \arxiv{arXiv:
  https://pubs.aip.org/aip/pfl/article-pdf/18/6/750/12272173/750\_1\_online.pdf}.

\bibitem[Munters {\em et~al.\/}(2016)Munters, Meneveau \& Meyers]{Munters2016}
{\sc \au{Munters, Wim}, \au{Meneveau, Charles} \& \au{Meyers, Johan}} \yr{2016}
   \at{Shifted periodic boundary conditions for simulations of wall-bounded
  turbulent flows}.  \jt{Physics of Fluids}  \bvol{28}~(2),  \pg{025112},
  \arxiv{arXiv:
  https://pubs.aip.org/aip/pof/article-pdf/doi/10.1063/1.4941912/15719452/025112\_1\_online.pdf}.

\bibitem[Naot {\em et~al.\/}(1973)Naot, Shavit \& Wolfshtein]{Naot1973}
{\sc \au{Naot, D.}, \au{Shavit, A.} \& \au{Wolfshtein, M.}} \yr{1973}
  \at{Two‐point correlation model and the redistribution of reynolds
  stresses}.  \jt{The Physics of Fluids}  \bvol{16}~(6),  \pg{738--743},
  \arxiv{arXiv:
  https://pubs.aip.org/aip/pfl/article-pdf/16/6/738/12386920/738\_1\_online.pdf}.

\bibitem[Niayifar \& Porté-Agel(2016)]{Niayifar2016}
{\sc \au{Niayifar, Amin} \& \au{Porté-Agel, Fernando}} \yr{2016}
  \at{Analytical modeling of wind farms: A new approach for power prediction}.
  \jt{Energies}  \bvol{9}~(9).

\bibitem[Pedersen {\em et~al.\/}(2022)Pedersen, Svensson, Poulsen \&
  Nygaard]{Pedersen2022}
{\sc \au{Pedersen, J~G}, \au{Svensson, E}, \au{Poulsen, L} \& \au{Nygaard,
  N~G}} \yr{2022}  \at{Turbulence optimized park model with gaussian wake
  profile}.  \jt{Journal of Physics: Conference Series}  \bvol{2265}~(2),
  \pg{022063}.

\bibitem[Pedersen {\em et~al.\/}(2023)Pedersen, Forsting, van~der Laan, Riva,
  Romàn, Risco, Friis-Møller, Quick, Christiansen, Rodrigues, Olsen \&
  Réthoré]{pyWake2023}
{\sc \au{Pedersen, Mads~M.}, \au{Forsting, Alexander~Meyer}, \au{van~der Laan,
  Paul}, \au{Riva, Riccardo}, \au{Romàn, Leonardo A.~Alcayaga}, \au{Risco,
  Javier~Criado}, \au{Friis-Møller, Mikkel}, \au{Quick, Julian},
  \au{Christiansen, Jens Peter~Schøler}, \au{Rodrigues, Rafael~Valotta},
  \au{Olsen, Bjarke~Tobias} \& \au{Réthoré, Pierre-Elouan}} \yr{2023}
  \at{Pywake 2.5.0: An open-source wind farm simulation tool}.  \jt{DTU Wind} .

\bibitem[Pe{\~n}a {\em et~al.\/}(2016)Pe{\~n}a, R{\'e}thor{\'e} \& van~der
  Laan]{pena2016application}
{\sc \au{Pe{\~n}a, Alfredo}, \au{R{\'e}thor{\'e}, Pierre-Elouan} \& \au{van~der
  Laan, M~Paul}} \yr{2016}  \at{On the application of the jensen wake model
  using a turbulence-dependent wake decay coefficient: the sexbierum case}.
  \jt{Wind Energy}  \bvol{19}~(4),  \pg{763--776}.

\bibitem[Perot \& Natu(2004)]{Perot2004}
{\sc \au{Perot, J.} \& \au{Natu, Sadbhaw}} \yr{2004}  \at{A model for the
  dissipation tensor in inhomogeneous and anisotropic turbulence}.  \jt{Physics
  of Fluids}  \bvol{16},  \pg{4053--4065}.

\bibitem[Pope(2000)]{Pope2000}
{\sc \au{Pope, Stephen~B.}} \yr{2000} {\em Turbulent Flows\/}.  \publ{Cambridge
  University Press}.

\bibitem[Rodier {\em et~al.\/}(2017)Rodier, Masson, Couvreux \&
  Paci]{Rodier2017}
{\sc \au{Rodier, Quentin}, \au{Masson, Valéry}, \au{Couvreux, Fleur} \&
  \au{Paci, Alexandre}} \yr{2017}  \at{Evaluation of a buoyancy and shear based
  mixing length for a turbulence scheme}.  \jt{Frontiers in Earth Science}
  \bvol{Volume 5 - 2017}.

\bibitem[{Rotta}(1951)]{Rotta1951}
{\sc \au{{Rotta}, J.}} \yr{1951}  \at{{Statistische Theorie nichthomogener
  Turbulenz}}.  \jt{Zeitschrift fur Physik}  \bvol{129}~(6),  \pg{547--572}.

\bibitem[Sarkar \& Speziale(1990)]{Sarkar1990}
{\sc \au{Sarkar, Sutanu} \& \au{Speziale, Charles~G.}} \yr{1990}  \at{A simple
  nonlinear model for the return to isotropy in turbulence}.  \jt{Physics of
  Fluids A: Fluid Dynamics}  \bvol{2}~(1),  \pg{84--93},  \arxiv{arXiv:
  https://pubs.aip.org/aip/pof/article-pdf/2/1/84/12712248/84\_1\_online.pdf}.

\bibitem[Schmidt {\em et~al.\/}(2023)Schmidt, Vollmer, Dörenkämper \&
  Stoevesandt]{Schmidt2023}
{\sc \au{Schmidt, Jonas}, \au{Vollmer, Lukas}, \au{Dörenkämper, Martin} \&
  \au{Stoevesandt, Bernhard}} \yr{2023}  \at{Foxes: Farm optimization and
  extended yield evaluation software}.  \jt{Journal of Open Source Software}
  \bvol{8}~(86),  \pg{5464}.

\bibitem[Schottenhamml {\em et~al.\/}(2022)Schottenhamml, Anciaux-Sedrakian,
  Blondel, Borras-Nadal, Joulin \& Rüde]{Schottenhamml2022}
{\sc \au{Schottenhamml, Helen}, \au{Anciaux-Sedrakian, Ani}, \au{Blondel,
  Frédéric}, \au{Borras-Nadal, Adria}, \au{Joulin, Pierre-Antoine} \&
  \au{Rüde, Ulrich}} \yr{2022}  \at{Evaluation of a lattice boltzmann-based
  wind-turbine actuator line model against a navier-stokes approach}.
  \jt{Journal of Physics: Conference Series}  \bvol{2265}~(2),  \pg{022027}.

\bibitem[Schottenhamml {\em et~al.\/}(2024)Schottenhamml, Anciaux~Sedrakian,
  Blondel, Köstler \& Rüde]{Schottenhamml2024}
{\sc \au{Schottenhamml, Helen}, \au{Anciaux~Sedrakian, Ani}, \au{Blondel,
  Frédéric}, \au{Köstler, Harald} \& \au{Rüde, Ulrich}} \yr{2024}
  \at{walberla-wind: A lattice-boltzmann-based high-performance flow solver for
  wind energy applications}.  \jt{Concurrency and Computation: Practice and
  Experience}  \bvol{36}~(16),  \pg{e8117},  \arxiv{arXiv:
  https://onlinelibrary.wiley.com/doi/pdf/10.1002/cpe.8117}.

\bibitem[Schreiber {\em et~al.\/}(2020)Schreiber, Balbaa \&
  Bottasso]{schreiber2020brief}
{\sc \au{Schreiber, Johannes}, \au{Balbaa, Amr} \& \au{Bottasso, Carlo~L}}
  \yr{2020}  \at{Brief communication: A double-gaussian wake model}.  \jt{Wind
  Energy Science}  \bvol{5}~(1),  \pg{237--244}.

\bibitem[Speziale {\em et~al.\/}(1992)Speziale, Abid \& Anderson]{Speziale1992}
{\sc \au{Speziale, Charles~G.}, \au{Abid, Ridha} \& \au{Anderson, E.~Clay}}
  \yr{1992}  \at{Critical evaluation of two-equation models for near-wall
  turbulence}.  \jt{AIAA Journal}  \bvol{30}~(2),  \pg{324--331},
  \arxiv{arXiv: https://doi.org/10.2514/3.10922}.

\bibitem[Speziale {\em et~al.\/}(1991)Speziale, Sarkar \& Gatski]{Speziale1991}
{\sc \au{Speziale, Charles~G.}, \au{Sarkar, Sutanu} \& \au{Gatski, Thomas~B.}}
  \yr{1991}  \at{Modelling the pressure–strain correlation of turbulence: an
  invariant dynamical systems approach}.  \jt{Journal of Fluid Mechanics}
  \bvol{227},  \pg{245–272}.

\bibitem[Spinelli {\em et~al.\/}(2023)Spinelli, Gericke, Masilamani \&
  Klimach]{Spinelli2023}
{\sc \au{Spinelli, Gregorio~Gerardo}, \au{Gericke, Jana}, \au{Masilamani,
  Kannan} \& \au{Klimach, Harald~G{\"u}nther}} \yr{2023}  \at{Key ingredients
  for wall-modeled les with the lattice boltzmann method: Systematic comparison
  of collision schemes, sgs models, and wall functions on simulation accuracy
  and efficiency for turbulent channel flow}.  \jt{Discrete and Continuous
  Dynamical Systems - Series S}  \pg{pp. 1--28}.

\bibitem[Stein \& Kaltenbach(2019)]{Stein2019}
{\sc \au{Stein, Victor~P.} \& \au{Kaltenbach, Hans-Jakob}} \yr{2019}
  \at{Non-equilibrium scaling applied to the wake evolution of a model scale
  wind turbine}.  \jt{Energies}  \bvol{12}~(14).

\bibitem[Stevens {\em et~al.\/}(2014)Stevens, Graham \& Meneveau]{Stevens2014}
{\sc \au{Stevens, Richard~J.A.M.}, \au{Graham, Jason} \& \au{Meneveau,
  Charles}} \yr{2014}  \at{A concurrent precursor inflow method for large eddy
  simulations and applications to finite length wind farms}.  \jt{Renewable
  Energy}  \bvol{68},  \pg{46--50}.

\bibitem[Stevens {\em et~al.\/}(2018)Stevens, Martínez-Tossas \&
  Meneveau]{Stevens2018}
{\sc \au{Stevens, Richard~J.A.M.}, \au{Martínez-Tossas, Luis~A.} \&
  \au{Meneveau, Charles}} \yr{2018}  \at{Comparison of wind farm large eddy
  simulations using actuator disk and actuator line models with wind tunnel
  experiments}.  \jt{Renewable Energy}  \bvol{116},  \pg{470--478}.

\bibitem[Stull(1988)]{Stull1988}
{\sc \au{Stull, Roland~B.}}, ed. \yr{1988} {\em An Introduction to Boundary
  Layer Meteorology\/}.  \publ{Springer Netherlands}.

\bibitem[Tennekes \& Lumley(1972)]{Tennekes1972}
{\sc \au{Tennekes, Henk} \& \au{Lumley, John~L.}} \yr{1972} {\em A First Course
  in Turbulence\/}.  \publ{Cambridge, Massachusetts: MIT Press}.

\bibitem[Tomas {\em et~al.\/}(2011)Tomas, Eiff \& Masson]{Tomas2011}
{\sc \au{Tomas, Séverine}, \au{Eiff, Olivier} \& \au{Masson, Valéry}}
  \yr{2011}  \at{Experimental investigation of turbulent momentum transfer in a
  neutral boundary layer over a rough surface}.  \jt{Boundary-Layer
  Meteorology}  \bvol{138},  \pg{385--411}.

\bibitem[Verstappen(2011)]{Verstappen2011}
{\sc \au{Verstappen, Roel}} \yr{2011}  \at{When does eddy viscosity damp
  subfilter scales sufficiently?}  \jt{Journal of Scientific Computing} .

\bibitem[Verstappen {\em et~al.\/}(2014)Verstappen, Rozema \&
  Bae]{Verstappen2014}
{\sc \au{Verstappen, Roel}, \au{Rozema, Wybe} \& \au{Bae, H.~Jane}} \yr{2014}
  Numerical scale separation in large-eddy simulation.  \bt{In {\em 15th
  Biennial Summer Program of the Center for Turbulence Research\/}}.

\bibitem[Wilcox(2008)]{Wilcox2008}
{\sc \au{Wilcox, David~C.}} \yr{2008}  \at{Formulation of the k-w turbulence
  model revisited}.  \jt{AIAA Journal}  \bvol{46}~(11),  \pg{2823--2838},
  \arxiv{arXiv: https://doi.org/10.2514/1.36541}.

\bibitem[Wu \& Porté-Agel(2011)]{Wu2011}
{\sc \au{Wu, Yu-Ting} \& \au{Porté-Agel, Fernando}} \yr{2011}  \at{Large-eddy
  simulation of wind-turbine wakes: Evaluation of turbine parametrisations}.
  \jt{Boundary-Layer Meteorology}  \bvol{138}.

\bibitem[Xie \& Archer(2015)]{Xie-Archer}
{\sc \au{Xie, Shengbai} \& \au{Archer, Cristina}} \yr{2015}
  \at{Self-similarity and turbulence characteristics of wind turbine wakes via
  large-eddy simulation}.  \jt{Wind Energy}  \bvol{18}~(10),  \pg{1815--1838},
  \arxiv{arXiv: https://onlinelibrary.wiley.com/doi/pdf/10.1002/we.1792}.

\end{thebibliography}

\end{document}